\input harvmac
\input epsf
\input amssym
%\draftmode
\baselineskip 13pt

%\pdfpageheight 0pt
%\pdfpagewidth 0pt

%%%%%%%%%ERIC's DEFS:
\def\cH{{\cal H}}

\def\sech{{\rm sech}}
\def\e{e^{-iy_1y_2}}

\def\L{\Lambda}

\def\hsl{hs[$\lambda$]}
\def\Hsl{HS[$\lambda$]}

\def\Zb{\overline{Z}}

\def\ah{\widehat{a}}

\def\Lr{\L_{\rho}}
\def\Lbr{\overline{\Lambda}_{\rho}}

\def\p{\partial}
\def\Wc{\cal{W}}
\def\Lc{{\cal L}}
\def\Oc{{\cal O}}
\def\half{{1\over 2}}
\def\rar{\rightarrow}
\def\mw{{\cal{W}}}
\def\ml{{\cal{L}}}
\def\a{\alpha}
\def\b{\beta}
\def\o{\omega}
\def\l{\lambda}
\def\t{\tau}
\def\eps{\epsilon}

\def\ep{e^{\rho}}
\def\ept{e^{2\rho}}
\def\emp{e^{-\rho}}
\def\zbar{\bar{z}}

\def\Ab{\overline{A}}
\def\mlb{\overline{\ml}}
\def\mwb{\overline{\mw}}
\def\ab{\overline{a}}

\def\zb{\overline{z}}
\def\taub{\overline{\tau}}

\def\Lcb{\overline{\cal L}}
\def\mub{\overline{\mu}}
\def\alphab{\overline{\alpha}}

\def\gt{\tilde{g}}

\def\vs{\vskip .1 in}
\def\bul{$\bullet$~}
\def\Ocb{\overline{\cal O}}
\def\phib{\overline{\phi}}
\def\tg{\tilde{g}}
\def\IZ{\Bbb{Z}}

%%%%%%%%%%%%%%%%%%%%%%%%%%%%%%%%%%%%%%%%%%%%%%
%%%%%%%%%%%%%%%%%%%%%%%%%%%
% some stuff needed for figures:
%%%%%%%%%%%%%%%%%%%%%%%%%%%%%%%%%%%%%%%%%%%%%%
%%%%%%%%%%%%%%%%%%%%%%%%%%%
\newcount\figno
\figno=0
\def\fig#1#2#3{
\par\begingroup\parindent=0pt\leftskip=1cm\rightskip=1cm\parindent=0pt
\baselineskip=11pt
\global\advance\figno by 1
\midinsert
\epsfxsize=#3
\centerline{\epsfbox{#2}}
\vskip -21pt
{\bf Fig.\ \the\figno: } #1\par
\endinsert\endgroup\par
}
\def\figlabel#1{\xdef#1{\the\figno}}
\def\encadremath#1{\vbox{\hrule\hbox{\vrule\kern8pt\vbox{\kern8pt
\hbox{$\displaystyle #1$}\kern8pt}
\kern8pt\vrule}\hrule}}
%%%%%%%%%%%%%%%%%%%%%%%%%%%%%%%%%%%%%%%%%%%%%%

\def\p{\partial}

\def\rt{\rightarrow}
\def\Oc{{\cal O}}

\def\zb{\overline{z}}

\def\taub{\overline{\tau}}

\def\eps{\epsilon}

\def\Ab{\overline{A}}

\def\zb{\overline{z}}
\def\Lc{{\cal L}}
\def\Lcb{\overline{\cal L}}

\def\Wc{{\cal W}}

\def\ab{\overline{a}}
\def\omb{\overline{\omega}}
\def\mub{\overline{\mu}}

\def\Hb{\overline{H}}
\def\omb{\overline{\omega}}

\def\gb{\overline{g}}

\def\hb{\overline{h}}

\def\gt{\tilde{g}}

\def\alphab{\overline{\alpha}}

\def\Lamb{\overline{\Lambda}}

\def\dtwo{\delta^{(2)}}
\def\sech{ {\rm sech} }
\def\gb{\overline{g}}

\def\cH{{\cal H}}
\def\IZ{\Bbb{Z}}

%\MukundaFV
\lref\MukundaFV{
  N.~Mukunda, E.~C.~G.~Sudarshan, J.~K.~Sharma and C.~L.~Mehta,
  ``Representations And Properties Of Parabose Oscillator Operators. I. Energy Position And Momentum Eigenstates,''
J.\ Math.\ Phys.\  {\bf 21}, 2386 (1980)..
}

%\MehtaBY
\lref\MehtaBY{
  C.~L.~Mehta, N.~Mukunda, J.~Sharma and G.~Sudarshan,
  ``Representations and properties of para-Bose oscillator operators II. Coherent states and the minimum uncertainty states,''
J.\ Math.\ Phys.\  {\bf 22}, 78 (1981)..
}

%\MehtaTJ
\lref\MehtaTJ{
  C.~L.~Mehta,
  ``Ordering of the Exponential of a Quadratic in Boson Operators. 1. Single Mode Case,''
J.\ Math.\ Phys.\  {\bf 18}, 404 (1977)..
}
%

%\AgarwalWC
\lref\AgarwalWC{
  G.~S.~Agarwal and E.~Wolf,
  ``Calculus for functions of noncommuting operators and general phase-space methods in quantum mechanics. i. mapping theorems and ordering of functions of noncommuting operators,''
Phys.\ Rev.\ D {\bf 2}, 2161 (1970)..
}

%\GaberdielKU
\lref\GaberdielKU{
  M.~R.~Gaberdiel and R.~Gopakumar,
  ``Triality in Minimal Model Holography,''
JHEP {\bf 1207}, 127 (2012).
[arXiv:1205.2472 [hep-th]].
%%CITATION = arXiv:1205.2472%%
}

%\DattaKM
\lref\DattaKM{
  S.~Datta and J.~R.~David,
  ``Supersymmetry of classical solutions in Chern-Simons higher spin supergravity,''
[arXiv:1208.3921 [hep-th]].
%%CITATION = arXiv:1208.3921%%
}

%%%%%%%%%%%%%%%%%%% REFS:

%\BalasubramanianRE
\lref\BalasubramanianRE{
  V.~Balasubramanian and P.~Kraus,
  ``A stress tensor for anti-de Sitter gravity,''
  Commun.\ Math.\ Phys.\  {\bf 208}, 413 (1999)
  [arXiv:hep-th/9902121].
  %%CITATION = CMPHA,208,413;%%
}

%\SezginPV
\lref\SezginPV{
  E.~Sezgin and P.~Sundell,
  ``An Exact solution of 4-D higher-spin gauge theory,''
Nucl.\ Phys.\ B {\bf 762}, 1 (2007).
[hep-th/0508158].
%%CITATION = hep-th/0508158%%
}

%\IazeollaWT
\lref\IazeollaWT{
  C.~Iazeolla, E.~Sezgin and P.~Sundell,
  ``Real forms of complex higher spin field equations and new exact solutions,''
Nucl.\ Phys.\ B {\bf 791}, 231 (2008).
[arXiv:0706.2983 [hep-th]].
%%CITATION = arXiv:0706.2983%%
}

%\DidenkoTD
\lref\DidenkoTD{
  V.~E.~Didenko and M.~A.~Vasiliev,
  ``Static BPS black hole in 4d higher-spin gauge theory,''
Phys.\ Lett.\ B {\bf 682}, 305 (2009).
[arXiv:0906.3898 [hep-th]].
%%CITATION = arXiv:0906.3898%%
}

%\IazeollaCB
\lref\IazeollaCB{
  C.~Iazeolla and P.~Sundell,
  ``Families of exact solutions to Vasiliev's 4D equations with spherical, cylindrical and biaxial symmetry,''
JHEP {\bf 1112}, 084 (2011).
[arXiv:1107.1217 [hep-th]].
%%CITATION = arXiv:1107.1217%%
}

  %\KlebanovJA
\lref\KlebanovJA{
  I.~R.~Klebanov and A.~M.~Polyakov,
  ``AdS dual of the critical O(N) vector model,''
Phys.\ Lett.\ B {\bf 550}, 213 (2002).
[hep-th/0210114].
%%CITATION = hep-th/0210114%%
}

%\ProkushkinBQ
\lref\ProkushkinBQ{
  S.~F.~Prokushkin and M.~A.~Vasiliev,
  ``Higher spin gauge interactions for massive matter fields in 3-D AdS space-time,''
Nucl.\ Phys.\ B {\bf 545}, 385 (1999).
[hep-th/9806236].
%%CITATION = hep-th/9806236%%
}

  %\SezginPT
\lref\SezginPT{
  E.~Sezgin and P.~Sundell,
  ``Holography in 4D (super) higher spin theories and a test via cubic scalar couplings,''
JHEP {\bf 0507}, 044 (2005).
[hep-th/0305040].
%%CITATION = hep-th/0305040%%
}

%\TanTJ
\lref\TanTJ{
  H.~-S.~Tan,
  ``Aspects of Three-dimensional Spin-4 Gravity,''
JHEP {\bf 1202}, 035 (2012).
[arXiv:1111.2834 [hep-th]].
%%CITATION = arXiv:1111.2834%%
}

%\GiombiVG
\lref\GiombiVG{
  S.~Giombi and X.~Yin,
  ``Higher Spins in AdS and Twistorial Holography,''
JHEP {\bf 1104}, 086 (2011).
[arXiv:1004.3736 [hep-th]].
%%CITATION = arXiv:1004.3736%%
}

%\GaberdielPZ
\lref\GaberdielPZ{
  M.~R.~Gaberdiel and R.~Gopakumar,
  ``An $AdS_3$ Dual for Minimal Model CFTs,''
Phys.\ Rev.\ D {\bf 83}, 066007 (2011).
[arXiv:1011.2986 [hep-th]].
%%CITATION = arXiv:1011.2986%%
}

%\GaberdielZW
\lref\GaberdielZW{
  M.~R.~Gaberdiel, R.~Gopakumar, T.~Hartman and S.~Raju,
  ``Partition Functions of Holographic Minimal Models,''
JHEP {\bf 1108}, 077 (2011).
[arXiv:1106.1897 [hep-th]].
%%CITATION = arXiv:1106.1897%%
}

%\GaberdielKU
\lref\GaberdielKU{
  M.~R.~Gaberdiel and R.~Gopakumar,
  ``Triality in Minimal Model Holography,''
[arXiv:1205.2472 [hep-th]].
%%CITATION = arXiv:1205.2472%%
}

%\BlencoweGJ
\lref\BlencoweGJ{
  M.~P.~Blencowe,
  ``A Consistent Interacting Massless Higher Spin Field Theory In D = (2+1),''
  Class.\ Quant.\ Grav.\  {\bf 6}, 443 (1989).
  %%CITATION = CQGRD,6,443;%%
}

%\BergshoeffNS
\lref\BergshoeffNS{
  E.~Bergshoeff, M.~P.~Blencowe and K.~S.~Stelle,
  ``Area Preserving Diffeomorphisms And Higher Spin Algebra,''
  Commun.\ Math.\ Phys.\  {\bf 128}, 213 (1990).
  %%CITATION = CMPHA,128,213;%%
}

%\GaberdielWB
\lref\GaberdielWB{
  M.~R.~Gaberdiel and T.~Hartman,
  ``Symmetries of Holographic Minimal Models,''
  arXiv:1101.2910 [hep-th].
  %%CITATION = ARXIV:1101.2910;%%
}

%\HenneauxXG
\lref\HenneauxXG{
  M.~Henneaux and S.~J.~Rey,
  ``Nonlinear W(infinity) Algebra as Asymptotic Symmetry of Three-Dimensional
  Higher Spin Anti-de Sitter Gravity,''
  JHEP {\bf 1012}, 007 (2010)
  [arXiv:1008.4579 [hep-th]].
  %%CITATION = JHEPA,1012,007;%%
}

%\CampoleoniZQ
\lref\CampoleoniZQ{
  A.~Campoleoni, S.~Fredenhagen, S.~Pfenninger and S.~Theisen,
  ``Asymptotic symmetries of three-dimensional gravity coupled to higher-spin
  fields,''
  JHEP {\bf 1011}, 007 (2010)
  [arXiv:1008.4744 [hep-th]].
  %%CITATION = JHEPA,1011,007;%%
}

%\AmmonUA
\lref\AmmonUA{
  M.~Ammon, P.~Kraus and E.~Perlmutter,
  ``Scalar fields and three-point functions in D=3 higher spin gravity,''
[arXiv:1111.3926 [hep-th]].
%%CITATION = arXiv:1111.3926%%
}

%\ChangMZ
\lref\ChangMZ{
  C.~-M.~Chang and X.~Yin,
  ``Higher Spin Gravity with Matter in $AdS_3$ and Its CFT Dual,''
[arXiv:1106.2580 [hep-th]].
%%CITATION = arXiv:1106.2580%%
}

%\CastroIW
\lref\CastroIW{
  A.~Castro, R.~Gopakumar, M.~Gutperle and J.~Raeymaekers,
  ``Conical Defects in Higher Spin Theories,''
JHEP {\bf 1202}, 096 (2012).

[arXiv:1111.3381 [hep-th]].
%%CITATION = arXiv:1111.3381%%
}

%\GutperleKF
\lref\GutperleKF{
  M.~Gutperle and P.~Kraus,
  ``Higher Spin Black Holes,''
JHEP {\bf 1105}, 022 (2011).
[arXiv:1103.4304 [hep-th]].
%%CITATION = arXiv:1103.4304%%
}

%\AmmonNK
\lref\AmmonNK{
  M.~Ammon, M.~Gutperle, P.~Kraus and E.~Perlmutter,
  ``Spacetime Geometry in Higher Spin Gravity,''
JHEP {\bf 1110}, 053 (2011).
[arXiv:1106.4788 [hep-th]].
%%CITATION = arXiv:1106.4788%%
}

%\KrausDS
\lref\KrausDS{
  P.~Kraus and E.~Perlmutter,
  ``Partition functions of higher spin black holes and their CFT duals,''
JHEP {\bf 1111}, 061 (2011).
[arXiv:1108.2567 [hep-th]].
%%CITATION = arXiv:1108.2567%%
}

%\GaberdielYB
\lref\GaberdielYB{
  M.~R.~Gaberdiel, T.~Hartman and K.~Jin,
  ``Higher Spin Black Holes from CFT,''
JHEP {\bf 1204}, 103 (2012).
[arXiv:1203.0015 [hep-th]].
%%CITATION = arXiv:1203.0015%%
}

%\BanadosUE
\lref\BanadosUE{
  M.~Banados, R.~Canto and S.~Theisen,
  ``The Action for higher spin black holes in three dimensions,''
[arXiv:1204.5105 [hep-th]].
%%CITATION = arXiv:1204.5105%%
}

%\HawkingSW
\lref\HawkingSW{
  S.~W.~Hawking,
  ``Particle Creation by Black Holes,''
Commun.\ Math.\ Phys.\  {\bf 43}, 199 (1975), [Erratum-ibid.\  {\bf 46}, 206 (1976)]..
}

%\MaldacenaKR
\lref\MaldacenaKR{
  J.~M.~Maldacena,
  ``Eternal black holes in anti-de Sitter,''
JHEP {\bf 0304}, 021 (2003).
[hep-th/0106112].
%%CITATION = hep-th/0106112%%
}

%\KrausIV
\lref\KrausIV{
  P.~Kraus, H.~Ooguri and S.~Shenker,
  ``Inside the horizon with AdS / CFT,''
Phys.\ Rev.\ D {\bf 67}, 124022 (2003).
[hep-th/0212277].
%%CITATION = hep-th/0212277%%
}

%\PopeSR
\lref\PopeSR{
  C.~N.~Pope, L.~J.~Romans and X.~Shen,
  ``W(infinity) And The Racah-wigner Algebra,''
Nucl.\ Phys.\ B {\bf 339}, 191 (1990)..
%%CITATION = USC-89-HEP040%%
}

%\CastroFM
\lref\CastroFM{
  A.~Castro, E.~Hijano, A.~Lepage-Jutier and A.~Maloney,
  ``Black Holes and Singularity Resolution in Higher Spin Gravity,''
JHEP {\bf 1201}, 031 (2012).
[arXiv:1110.4117 [hep-th]].
%%CITATION = arXiv:1110.4117%%
}

%\IazeollaNF
\lref\IazeollaNF{
  C.~Iazeolla and P.~Sundell,
  ``Biaxially symmetric solutions to 4D higher-spin gravity,''
[arXiv:1208.4077 [hep-th]].
%%CITATION = arXiv:1208.4077%%
}

%\TanXI
\lref\TanXI{
  H.~S.~Tan,
  ``Exploring Three-dimensional Higher-Spin Supergravity based on sl(N |N - 1) Chern-Simons theories,''
[arXiv:1208.2277 [hep-th]].
%%CITATION = arXiv:1208.2277%%
}

%\CampoleoniHP
\lref\CampoleoniHP{
  A.~Campoleoni, S.~Fredenhagen, S.~Pfenninger and S.~Theisen,
  ``Towards metric-like higher-spin gauge theories in three dimensions,''
[arXiv:1208.1851 [hep-th]].
%%CITATION = arXiv:1208.1851%%
}

%\PerezCF
\lref\PerezCF{
  A.~Perez, D.~Tempo and R.~Troncoso,
  ``Higher spin gravity in 3D: black holes, global charges and thermodynamics,''
[arXiv:1207.2844 [hep-th]].
%%CITATION = arXiv:1207.2844%%
}

%\BalasubramanianDE
\lref\BalasubramanianDE{
  V.~Balasubramanian, P.~Kraus, A.~E.~Lawrence and S.~P.~Trivedi,
  ``Holographic probes of anti-de Sitter space-times,''
Phys.\ Rev.\ D {\bf 59}, 104021 (1999).
[hep-th/9808017].
%%CITATION = hep-th/9808017%%
}

%\BakasRY
\lref\BakasRY{
  I.~Bakas and E.~Kiritsis,
  ``BOSONIC REALIZATION OF A UNIVERSAL W ALGEBRA AND Z(infinity) PARAFERMIONS,''
Nucl.\ Phys.\ B {\bf 343}, 185 (1990), [Erratum-ibid.\ B {\bf 350}, 512 (1991)]..
%%CITATION = LBL-28714%%
}

%\AmmonWC
\lref\AmmonWC{
  M.~Ammon, M.~Gutperle, P.~Kraus and E.~Perlmutter,
  ``Black holes in three dimensional higher spin gravity: A review,''
[arXiv:1208.5182 [hep-th]].
%%CITATION = arXiv:1208.5182%%
}

%\GaberdielUJ
\lref\GaberdielUJ{
  M.~R.~Gaberdiel and R.~Gopakumar,
  ``Minimal Model Holography,''
[arXiv:1207.6697 [hep-th]].
%%CITATION = arXiv:1207.6697%%
}

%%%%%%%%%%%%%%%%%%%%%%%

\Title{\vbox{\baselineskip14pt
%\hbox{hep-th/0508218}
%\hbox{UCLA-05-TEP-XX} \hbox{MCTP-XX-XX}
}} {\vbox{\centerline {Probing higher spin black holes}}}
%\medskip\vbox{\centerline {and their CFT duals}}}}
\centerline{Per
Kraus and Eric Perlmutter\foot{pkraus@ucla.edu, ejperlmutter@gmail.com}}
\bigskip
\centerline{\it{Department of Physics and Astronomy}}
\centerline{${}$\it{University of California, Los Angeles, CA 90095, USA}}

\baselineskip14pt

\vskip .3in

\centerline{\bf Abstract}
\vskip.2cm
We study the propagation of scalar fields on various backgrounds in three dimensional higher spin gravity. Our main emphasis is on obtaining the bulk-boundary propagator, which can be efficiently computed using group theory and higher spin gauge symmetry and from which we can extract scalar two-point functions in the dual CFT. As an illustration, we obtain a simple closed form expression for the propagator in a particular spin-3 deformation of AdS$_3$.    In the case of higher spin black holes, we prove on general grounds that the propagator respects an imaginary time periodicity consistent with the thermal nature of the black hole; in doing so we make progress in understanding the group exponentiation of the higher spin Lie algebra \hsl, and its center.  We also explicitly compute the propagator in the black hole background at first order in the higher spin charge.  Evaluated on the Lorentzian section, the result is consistent with an interpretation in which the black hole has two causally disconnected boundary components, as is the case for the BTZ black hole.

%%%
\Date{September  2012}
%%%%%%%%%%%%%%%%%%%%%%%%%%%%%%%%%%%%%%%%%%%%%%
%%%%%%%%%%%%%%%%%%%%%%%%%%%
% Main text begins here
%%%%%%%%%%%%%%%%%%%%%%%%%%%%%%%%%%%%%%%%%%%%%%
%%%%%%%%%%%%%%%%%%%%%%%%%%%
\baselineskip13pt

\listtoc\writetoc
\newsec{Introduction}

Due to its appearance on the bulk side of  soluble examples of the AdS/CFT correspondence \refs{\KlebanovJA,\SezginPT,\GaberdielPZ,\GaberdielUJ}, higher spin gravity provides a promising arena for tackling fundamental problems in quantum gravity. Black hole physics is one obvious target for such investigations. In three dimensions it is by now understood how to construct black holes that generalize the BTZ solution to carry higher spin charge \refs{\GutperleKF,\AmmonNK,\KrausDS,\CastroFM,\TanTJ,\BanadosUE,\PerezCF,\TanXI,\DattaKM,\AmmonWC}.  The entropy of these black holes has been computed \refs{\GutperleKF,\KrausDS}, and matches that of the corresponding dual CFT \refs{\KrausDS,\GaberdielYB}.  Much less is known in four dimensions; see \refs{\SezginPV,\IazeollaWT,\DidenkoTD,\IazeollaCB,\IazeollaNF} for discussion of classical solutions.

In this paper we focus on further understanding the physical properties of three dimensional higher spin black holes.  A special property of the three dimensional theory \refs{\BlencoweGJ,\BergshoeffNS,\ProkushkinBQ,\HenneauxXG,\CampoleoniZQ} is that it admits a consistent truncation to a sector involving just the higher spin gauge fields without ``matter"; this sector is a Chern-Simons theory for two gauge fields ($A,\Ab)$, and is topological since the gauge fields have no local degrees of freedom.  The higher spin black holes that have been found lie entirely in this topological sector, which explains why they are relatively tractable to study.

Of course, what accounts for much of the interest in black holes is the effect of their curved geometry on propagating matter.   There exists a fully explicit, though complicated, set of equations governing the interaction of propagating scalar fields with the higher spin gauge fields \ProkushkinBQ.  Our goal here is to study the propagation of free scalar fields on a variety of higher spin backgrounds, including black holes.  We are mainly interested in computing the bulk-boundary propagator, from which the boundary two-point function of scalar operators can be extracted.  This problem already poses a variety of interesting challenges, of both the technical and conceptual sort, which we now discuss.

While on a pure AdS background the scalar field obeys an ordinary Klein-Gordon equation, on a background with higher spin fields present the scalar wave equation becomes a complicated higher order, typically nonlocal, differential equation.  Writing down this equation explicitly is already challenging (see \AmmonUA\ for some examples), much less solving it.  Fortunately, there exists an elegant procedure by which solutions can be written down directly.  Since the higher spin sector is topological, any background solution can be obtained from vanishing gauge  fields by a gauge transformation (setting aside global issues). Note that there is no notion of spacetime for vanishing gauge fields --- which we call ``$A=0$ gauge'' for short --- and so the scalar master field is described fully  by an element of the gauge algebra. Acting on this scalar master field with the gauge transformation that generates the desired gauge fields, one obtains the physical bulk-boundary propagator. This approach has been used previously to compute correlators in AdS \refs{\GiombiVG,\ChangMZ,\AmmonUA}.

The central question is {\it which} element of the gauge algebra should be taken to be the $A=0$ gauge scalar master field. This, in turn, is a question of what boundary conditions one wishes to enforce on the solution to the scalar wave equation. We argue that the $A=0$ gauge scalar master field corresponding to the bulk-boundary propagator takes the same simple form for all backgrounds; it is fully fixed by the higher spin symmetry. Support for this is provided by comparing to the properties of the dual scalar primary in the boundary CFT \refs{\GaberdielPZ,\AmmonUA}.

One of our auxiliary goals is to develop the technology required to carry out these computations efficiently.  The relevant gauge algebra is the infinite dimensional Lie algebra \hsl\ $\oplus$ \hsl, and the general expression for the propagator in an arbitrary higher spin background is given by a trace of a product of group elements.  Due to the complicated nature of the \hsl\ structure constants, these traces are difficult to evaluate.  The situation simplifies at the special value $\lambda = 1/2$: at this value of $\lambda$ the associative product underlying the \hsl\ Lie algebra is equivalent to the Moyal product acting on functions of a pair of spinor variables.

As an illustrative warmup example in the use of these techniques, we show how to derive the propagator in the BTZ black hole background.  From this propagator one reads off the scalar two-point function in the boundary theory.  As is well known, this is a thermal two-point function, periodic in imaginary time, a property that is inherited from the Euclidean BTZ metric.  Another well known fact is that the Lorentzian BTZ solution has two disconnected AdS$_3$ boundaries.  On the CFT side, the BTZ solution appears as a state in the tensor product of two CFTs, one associated to each boundary component \refs{\BalasubramanianDE,\MaldacenaKR}.  A thermal density matrix arises upon tracing over one factor.  One can also compute a ``mixed" two-point function, with one operator from each CFT component.   This is obtained from the bulk by taking the two arguments of the bulk-boundary propagator to live on distinct boundary components.

Interesting new features arise once we generalize to black holes carrying higher spin charge. As has been elaborated on in detail in previous work  \refs{\GutperleKF,\AmmonNK,\KrausDS}, due to the enlarged gauge symmetry of the higher spin theory, such basic features as the causal structure of the metric can be changed by a gauge transformation.  In fact, the gauge that is most convenient for constructing solutions and studying their properties yields a metric that is not that of a black hole, but rather a static, traversable wormhole, with no event horizon.  However, the metric is not the full story when higher spin fields are present, as the latter influence the propagation of signals and hence the causal structure.  For the case of black holes in spin-3 gravity, it was conjectured in \GutperleKF\ and established in \AmmonNK\ that there exists a spin-3 gauge transformation that takes the metric to that of a black hole.  Furthermore, the thermodynamics of the solution are consistent with a black hole interpretation.

To study the physical causal structure of a solution we need to probe it with dynamical matter, which is what we do here using scalar fields.  It again proves most convenient to compute the scalar propagator in the ``wormhole" gauge.  As emphasized above, in this gauge imaginary time can naively be compactified with any desired period, since the corresponding circle never shrinks to zero size metrically.  Nonetheless, we establish on general grounds that the scalar propagator has a specific periodicity, which precisely matches the periodicity assigned to the black hole in \KrausDS.  The logic used in \refs{\GutperleKF,\AmmonNK,\KrausDS} to assign the periodicity was based on demanding a well-behaved gauge holonomy around the thermal circle.   It is gratifying that scalar fields really do appear to ``see" a black hole solution in a precise sense, even though none is manifestly present in the gauge employed.
 
The black hole interpretation is further supported by explicit computation of the scalar propagator in the black hole background of \KrausDS, to first order in the higher spin charge.  Our result gives additional credence to the notion that the two boundaries of the Lorentzian black hole are causally disconnected, just as in BTZ. In particular, we can compute the mixed two-point function and observe that it is nonsingular.  Had the two boundaries instead been causally connected, we might expect there to be singularities on the ``lightcone".\foot{Here we are basing our intuition on the fact that  propagators computed from the curved spacetime Klein-Gordon equation are nonsingular at spacelike separation.  We therefore attribute singularities to the existence of causal curves connecting the two points. }  One noteworthy difference as compared to BTZ is the following.  For BTZ, the mixed correlator can be obtained from the single sided correlator by shifting imaginary time by half a period; one can see this in Kruskal type coordinates by noting that a half shift of imaginary time is equivalent to a reflection of the radial coordinate about the horizon \KrausIV. For the higher spin black hole this simple relation no longer holds.

It is also of interest to consider cases in which the scalar propagator can be computed exactly to all orders in the higher spin fields. We analyze such an example, a certain zero temperature limit of the black hole which is a spin-3 deformation of AdS$_3$.  The scalar two-point function extracted from this computation has modified short distance behavior, consistent with the addition of an irrelevant dimension-(3,0) operator to the CFT action.

Finally, there is a technical result that we wish to highlight which should have broader applications to higher spin gravity. Our analysis of the thermal periodicity of the scalar propagator demands a more fundamental understanding of black hole holonomy, defined as a path ordered  exponential of the \hsl\ gauge fields. To this point, little has been understood about the group exponentiation of \hsl, which we denote \Hsl.  Thermal periodicity requires that the holonomy lie in the center of \Hsl, and we show that this is true for both the BTZ and higher spin black hole. At $\l = 1/2$ we can be more explicit and establish that the central element is nontrivial, instead given by a delta function of the two-dimensional spinors appearing in the Moyal product description of \hsl.  We show that the center of HS$[{1\over 2}]$ contains at least a   $\IZ_4$ subgroup. In view of the discovery of smooth conical defects of SL(N) gravity \CastroIW, our results may be helpful in finding such solutions for non-integer $\l$.

The paper is organized as follows. Section 2 introduces the main idea for computing scalar bulk-boundary propagators in general spacetimes, including an explicit calculation of the propagator in AdS. Section 3 applies this methodology to the BTZ black hole, shows that the BTZ thermal holonomy lies in the center of \Hsl, and establishes the thermal periodicity of scalar propagators in general black hole backgrounds with the same holonomy as BTZ. In Section 4 we compute the scalar propagators in the higher spin backgrounds described above; this reveals the thermal nature of the black hole in a variety of ways.  Section 5 discusses the existence of a $\IZ_4$ subgroup of the center of HS[$\half$], laying groundwork for the pursuit of conical defects in hs[$\half$] gravity. We conclude in section 6.

\newsec{Scalar propagators from $A=0$ gauge}

In this section we discuss some general issues regarding the computation of scalar field bulk-boundary propagators in higher spin backgrounds.   Our main example in this section is the simplest case of a scalar propagating in AdS$_3$ with no higher spin fields turned on.

\subsec{General aspects}

The scalar field is contained in the master field $C$, which obeys the following linearized equation
\eqn\aa{ dC + A\star C - C\star \Ab =0~.}
The full Vasiliev theory contains a specific set of interaction terms, which modify \aa\ at {\cal O}$(C^2)$ and above (we refer to  \AmmonUA\ for more details, as well as to further references to the original literature of Vasiliev and collaborators).  The linearized equation \aa\ is sufficient to compute AdS/CFT correlation functions involving at most two scalar operators.

In \aa, $A$ and $\Ab$ are independent elements of the Lie algebra \hsl. The generators of \hsl\ are denoted as $V^s_m$, with $s=2, 3, 4, \ldots$, and $m = -(s-1), -(s-2), \ldots, s-1$.  The generators multiply according to the associative ``lone star product"  \PopeSR\ denoted by $\star$.   It is also useful to introduce the identity element $V^1_0$ which commutes with all elements of \hsl.  The master fields $C$, $A$, and $\Ab$ can then be expanded in generators as
\eqn\abb{\eqalign{ C &=\sum_{s=1}^\infty \sum_{|m|<s} C^s_m(x^\mu) V^s_m~, \cr
A &=\sum_{s=2}^\infty \sum_{|m|<s} A^s_m(x^\mu) V^s_m~, \cr
 \Ab &=\sum_{s=2}^\infty \sum_{|m|<s} \Ab^s_m(x^\mu) V^s_m~,}}
where $x^\mu$ denote the spacetime coordinates. See Appendix B for rudiments of \hsl\ and further references.

We define a formal ``trace" operation $\Tr$ by picking out the coefficient of $V^1_0$.  By this definition, $\Tr (C) = C^1_0$, while $A$ and $\Ab$ are traceless.    Up to a proportionality constant, the lowest component $C^1_0$ is identified with the physical scalar field, while the other components of $C$ are auxiliary fields that are related to the lowest component by the field equation \aa.

In \aa, $A$ and $\Ab$ are flat connections:
\eqn\ac{ dA + A\wedge \star A = 0~,\quad  d\Ab + \Ab\wedge \star \Ab = 0~.}
Indeed, by applying $d$ we see that \aa\ is inconsistent for general (i.e. non-flat) background connections.

The scalar equation is invariant under infinitesimal \hsl\ $\oplus$ \hsl\ gauge transformations,
\eqn\adda{ \eqalign{ A & \rt A + d\Lambda +[A,\Lambda]_\star \cr
  \Ab & \rt \Ab + d\Lamb +[\Ab,\Lamb]_\star \cr
C & \rt C + C\star \Lamb -\Lambda \star C }}
and their finite versions
\eqn\ae{\eqalign{A & \rt e^{-\Lambda} \star(d+A)\star e^{\Lambda} \cr
\Ab & \rt e^{-\Lamb}\star (d+\Ab)\star e^{\Lamb} \cr
C & \rt e^{-\Lambda}\star C\star e^{\Lamb}~.  }}
We note that the exponentials are defined by using $\star$ products in their series expansion.

As reviewed in \AmmonUA, evaluated in AdS$_3$ the scalar equation \aa\ boils down to a Klein-Gordon equation for a scalar field $\Phi$ of mass $m^2 = \lambda^2-1$.\foot{Note that we are setting the AdS$_3$ radius to unity.}  As noted above, this scalar field is given by the lowest component of the master field $C$:
\eqn\aea{\Phi = \Tr [C] = C^1_0}

\subsec{Moyal product at $\lambda =1/2$}

For the special value $\lambda=1/2$ the $\star$-product can be represented by the Moyal product for a pair of spinor variables $y_{1,2}$. For a pair of functions $f(y)$ and $g(y)$, the Moyal product in differential form is
\eqn\aea{(f* g)(y) = \exp \left[ i\eps_{\alpha \beta} {\p \over \p y_{\alpha}}{\p \over \p y'_\beta}\right] f(y) g(y')\Big|_{y'=y}  }
and its integral form is
\eqn\aeb{ (f * g)(y) = {1\over 4\pi^2} \int\! d^2 u d^2 v f(y+u) g(y+v) e^{i uv}}
with $uv=u_\alpha v^\alpha = \eps^{\alpha\beta}u_\alpha v_\beta$, and $\eps_{12}=\eps^{12}=1$.  The integral is defined so that $1 * 1 =1$.  An important fact about the Moyal product is that symmetrized star products are equivalent to ordinary products,
\eqn\aec{ (y_1+y_2)^{*n} = (y_1 + y_2)^n~,}
as is easily proven by induction.

To make contact with the $\star$-product, write
\eqn\aed{ V^s_m = \left({-i \over 4}\right)^{s-1} y_1^{s+m-1}y_2^{s-m-1}~.}
Moyal multiplication of the spinor version of the $V^s_m$ is then isomorphic to $\star$-product multiplication  at $\lambda=1/2$.

The trace operation in spinor language is
\eqn\aeda{ \Tr [f(y)] = f(0)~.}

\subsec{AdS scalar propagator}

We now discuss some general issues involving the computation of the scalar propagator.  We write the metric of Euclidean signature AdS$_3$ as
\eqn\af{ds^2 = d\rho^2  + e^{2\rho}dzd\zb~.}
For $0 \leq \lambda \leq 1$ the scalar field admits both standard and alternate quantizations.    The bulk-boundary propagator obeys the following boundary conditions as $\rho \rt \infty$:
\eqn\ag{G_{\pm}(\rho,{\bf x};{\bf x'}) \sim \left[e^{-(1\mp\l)\rho}\delta^{(2)}( {\bf x}-{\bf x'})+ \ldots \right] + \left[ e^{-(1\pm\l)\rho}\phi_1({\bf x-x'})+\ldots\right] ~. }
Subscripts $+$ and $-$ denote standard and alternate quantizations, respectively, and $\phi_1({\bf x-x'})$ is the scalar vev term.

In the coordinates \af\ we have
\eqn\ah{ G_\pm(\rho,{\bf{x}}; {\bf{x'}} ) = \pm {\lambda \over \pi} \left({e^{-\rho} \over e^{-2\rho}+|z-z'|^2}\right)^{1\pm \lambda}~.  }

To obtain this result from \aa\ we first write down the connections representing the background \af.  In general, it is convenient to choose a gauge such that
\eqn\aha{\eqalign{ A(\rho,z,\zb) &= a_z(z,\zb)dz+a_{\zb}(z,\zb)d\zb +V^2_0 d\rho\cr
& = b^{-1}ab +b^{-1}db \cr
 \Ab(\rho,z,\zb) &= \ab_z(z,\zb)dz+\ab_{\zb}(z,\zb)d\zb -V^2_0 d\rho\cr
& = b^{1}\ab b^{-1} +bdb^{-1}~,}}
with $b= e^{\rho V^2_0}$. Flatness of $(A,\Ab)$  implies flatness of $(a,\ab)$.    AdS in the coordinates of \af\ is then represented by
\eqn\ai{\eqalign{a_{AdS} &= V_1^2 dz  \cr
\ab_{AdS} & =  V^2_{-1} d\zb~. }}
To verify this,  recall that the generalized vielbein $e$ and spin connection $\omega$ are defined as $A = \omega+e$, $\Ab =\omega -e$, and  $g_{\mu\nu} \propto \Tr (e_\mu e_\nu)$ with some convenient normalization.

Throughout the present work $a$ and $\ab$ will be independent of $z$ and $\zb$, so we can  write $a=d\Lambda$ and $\ab = d\Lamb$ with
\eqn\aia{\eqalign{\Lambda &= a_{\mu}x^\mu \cr \Lamb &= \ab_{\mu}x^\mu~.}}
Writing
\eqn\aib{\eqalign{ g&= e^\Lambda \star b~,\cr
\gb &= e^{\Lamb} \star b^{-1}~,}}
we have
\eqn\aic{\eqalign{ A & = g^{-1} \star dg \cr
\Ab & = \gb^{-1}\star d \overline{g}~,}}
which makes it manifest that the connections are (locally) pure gauge.
Before proceeding further we introduce one more piece of notation, writing
\eqn\aid{\eqalign{ \Lambda_\rho &= b^{-1} \star \Lambda \star b \cr
\Lamb_\rho &= b \star \Lamb \star  b^{-1}~.}}
$\Lambda_\rho$ is obtained from $\Lambda$ by the replacement $V^s_m \rt e^{m\rho} V^s_m$; similarly, $\Lamb_\rho$ is obtained from $\Lamb$ by the replacement $V^s_m \rt e^{-m\rho} V^s_m$.

Using the fact that the connections are locally pure gauge, we can now describe a simple method for obtaining the scalar propagator.  Since the scalar equation \aa\ is gauge covariant, we can first solve it in the gauge $A=\Ab=0$ and then act with a gauge transformation to obtain the solution in AdS.   It is of course trivial to write down solutions of \aa\ if $A=\Ab=0$: simply take any linear combination of $V^s_m$ with constant coefficients. Gauge transformation will then produce solutions of the Klein-Gordon equation in AdS.  Explicitly, we use $c$ to denote the scalar master field $C$ in this gauge, which we henceforth call ``$A=0$ gauge'' for short.    This  master  field in AdS is then
\eqn\al{\eqalign{C &= g^{-1}\star c \star \gb\cr
& =e^{-\Lambda_\rho} \star b^{-1} \star c \star b^{-1} \star e^{\overline{\Lambda}_\rho}   ~.}}
This approach is similar in spirit to that of \GiombiVG\ in which the authors compute boundary three-point functions in 4d Vasiliev theory (there, the spacetime-independent gauge is called the ``$W=0$ gauge'').

The real question is which choices for $c$ yield the solutions in AdS obeying the boundary conditions \ag.   The answer to this question turns out to be much simpler at $\lambda=1/2$ where we can use the Moyal product, so we first focus on that case.

\subsec{AdS scalar propagator at $\lambda=1/2$}

We claim that taking
\eqn\am{ c_- = e^{-iy_1 y_2}~,\quad c_+ =\left({1\over 2i}\right) y_1 * e^{-i y_1 y_2}* y_2}
yields
\eqn\an{ G_{\pm}(\rho,{\bf x};{\bf 0}) = \pm{1\over 2\pi} \Tr[C_{\pm}]}
where $G_\pm$ refer to the AdS propagators \ah.
The origin of these expressions will be explained in the next subsection; here we just verify \an\ by direct computation.

We first note that $c_\pm$ are eigenfunctions of $V^2_0$ acting from the left or right; indeed, for arbitrary $n$ we have
\eqn\ao{ V^2_0 *  y_1^n * e^{-iy_1 y_2}*y_2^n =y_1^n * e^{-iy_1 y_2}*y_2^n* V^2_0= -\left({1\over 4} + {n\over 2}\right) y_1^n * e^{-iy_1 y_2}*y_2^n~. }
This implies
\eqn\ap{\eqalign{ C_\pm
& = e^{(n_\pm +{1\over 2})\rho}  \Big[ e^{-\Lambda_\rho} * c_\pm * e^{\Lamb_\rho}\Big] }}
with $n_-=0$ and $n_+=1$; the trace gives the scalar field, cf. \aea.  In terms of the spinor variables we have
\eqn\aq{ \Lambda_\rho = -{i\over 4}z e^{\rho} y_1^2~,\quad  \Lamb_\rho = -{i\over 4}\zb e^{-\rho} y_2^2~.}
Using results in Appendix A.2 we have
\eqn\ar{  e^{-\Lambda_\rho} * e^{-iy_1 y_2} * e^{\Lamb_\rho} ={1\over\sqrt{|L|}} e^{{1\over 2}y^T S y}  }
with
\eqn\as{ |L| = e^{2\rho}z\zb+1 ~,\quad S = {i\over e^{2\rho} z\zb+1}\left(\matrix{2 e^\rho z & e^{2\rho}z\zb-1 \cr
e^{2\rho}z\zb-1 & -2 e^{\rho} \zb }\right)~.}
This immediately yields
\eqn\at{ \Tr \big[C_-\big]= {e^{{1\over 2}\rho} \over (e^{2\rho}z\zb +1)^{1/2} }  }
from which we see that \an\ agrees with \ah\ for $\lambda =1/2$.

Slightly more work is required for for the case of standard quantization.  In this case we have
\eqn\au{\eqalign{ \Tr\big[C_+\big]& ={e^{{3\over 2}\rho}\over2i \sqrt{e^{2\rho}z\zb +1}}\Tr\Big[ y_1 * e^{{1\over 2}y^T S y} * y_2 \Big] \cr
& =  {e^{{3\over 2}\rho}\over 2i\sqrt{e^{2\rho}z\zb +1}}(i-S_{12}) \cr
& =  {e^{{3\over 2}\rho}\over (e^{2\rho}z\zb +1)^{3/2}}  }}
which verifies \an.

For the remainder of the paper, we adhere to the notation $\Phi_{\pm}$ rather than $\Tr[C_{\pm}]$, and take the liberty of referring to $\Phi_{\pm}$ as ``propagators,'' with the convention as in \an\ that the actual propagators including the correct overall normalization are given as
\eqn\tsa{G_{\pm}(\rho,{\bf x};{\bf 0}) = \pm{\l\over\pi}\Phi_{\pm}~.}
%}
%

\subsec{AdS scalar propagator as a \hsl\ eigenvalue problem}
We just showed that at $\l=1/2$, the $A=0$ gauge master fields $c_{\pm}$ in \am\ giving rise to the AdS propagators obeying the boundary conditions \ag\ are eigenfunctions of $V^2_0$. This turns out to be a key observation. We now argue that for generic $\l$, taking $c_{\pm}$ to be highest weight states of \hsl\ leads, upon gauge transformation to AdS, to the correct bulk-boundary scalar propagators $\Phi_{\pm}$ with delta-function boundary conditions.

For motivation, we turn to CFT. Recall that the bulk scalar is dual to a scalar primary $\Oc$ in a CFT with  $\mw_{\infty}[\l]$ symmetry, which is the asymptotic symmetry algebra of  \hsl\ gravity in AdS$_3$ \GaberdielWB. This symmetry algebra has conserved spin-$s$ currents $J^{(s)}$, with $s=2,3,\ldots$, and mode expansions
\eqn\eijc{J^{(s)}(z) = -{1\over 2\pi}\sum_{m\in Z}{J^{(s)}_m\over z^{s+m}}~.}
 The OPE of $J^{(s)}$ and $\Oc_{\pm}$ has the following leading singularity,
\eqn\eijb{J^{(s)}(z)\Oc_{\pm}(0) \sim {A_{\pm}(s)\over z^s}\Oc_{\pm}(0)+\ldots}
where
\eqn\eij{ A_{\pm}(s) = {(-1)^{s+1}\over 2\pi}{\Gamma(s)^2\over \Gamma(2s-1)}{\Gamma(s\pm \l)\over \Gamma(1\pm \l)}~.}
This coefficient was derived in \AmmonUA, and we will re-derive it below in an intuitive way. The modes with $m<|s|$ are the `wedge modes' which generate the infinite-dimensional Lie algebra \hsl\ of the bulk:
\eqn\eijca{J^{(s)}_m = V^s_m~, ~~ |m|<s~.}

The CFT primary state $|\Oc_{\pm}\rangle$ is a highest weight state and an eigenstate of the spin-$s$ zero modes $V^s_0$,
\eqn\eih{\eqalign{V^s_0|\Oc_{\pm}\rangle &= -2\pi A_{\pm}(s)|\Oc_{\pm}\rangle\cr
V^s_{m>0}|\Oc_{\pm}\rangle&=0~.}}
Returning to the bulk Vasiliev theory, our first claim is that the $A=0$ gauge master fields $c_{\pm}$ which lead to scalar propagators in AdS with delta-function boundary conditions are highest weight states of \hsl, and hence obey the analogous equations
\eqn\eik{\eqalign{V^s_0\star c_{\pm} &= x_{\pm}(s)c_{\pm}\cr
V^s_{m<0}\star c_{\pm} &= 0~.}}
The $x_{\pm}(s)$ are taken to be unknown coefficients at this point, but we will soon find $x_{\pm}(s)=2\pi A_{\pm}(s)$ as expected up to a sign convention.\foot{Signs differ between \eih\ and \eik\ on account of convenient bulk vs. CFT convention choices. This could be changed by replacing left- by right-multiplication in \eik. The lack of a minus sign in the definition of $x_{\pm}(s)$ also differs from the CFT equations but is correct in the bulk; see equation (2.36), for example.}

The resulting $c_{\pm}$ match previous results obtained from AdS master fields \ChangMZ. Similar statements can be made about the conformal descendants.

Before solving equations \eik, a quick way to see that $c_{\pm}$ can be an eigenfunction of $V^2_0$ --- and hence of all (commuting) zero modes, which can be written as linear combinations of powers of $V^2_0$ and the identity, $V^1_0$ --- is as follows. In AdS, at the origin on the boundary $z=\zb=0$ the propagator is (up to overall normalization)
\eqn\sca{\eqalign{\Phi^{AdS}_{\pm}(z=\zb=0) &= e^{(1\pm\l)\rho}\cr
&=\Tr[e^{-\Lr^{AdS}}\star b^{-1} \star c_{\pm}\star b^{-1} \star e^{\Lbr^{AdS}}]\Big|_{z=\zb=0}\cr
&= \Tr[b^{-1} \star c_{\pm}\star b^{-1}]}}
where we use $\Lr(z=\zb=0)=\Lbr(z=\zb=0)=0$.   This is compatible with $V^2_0 \star c_\pm = -{1\over 2} (1\pm \lambda) c_\pm$.

Let us solve \eik\ and drop the $\pm$ subscript for now. First, it is clear that $c$ must only include zero mode generators because $(z,\zb)$ only appear in the AdS scalar propagator in the combination $z\zb$, cf. (2.24) and (2.25). Accordingly, expand $c$ in generators as
\eqn\eil{c = \sum_{t=1}^{\infty}c_{t}V^t_0}
and use the overall normalization of $c$ to set $c_{1}=1$. Taking the trace of the first equation, we have
\eqn\eij{c_{t} = {x(t)\over \Tr(V^t_0V^t_0)}~.}
The trace is known (see Appendix B), so we only need to find $x(t)$.

Taking $s=2,m=-1$ in \eik, we extract the spin-$t$ components of each equation to obtain two distinct relations among the coefficients $(c_{t-1},c_t,c_{t+1})$ in terms of some \hsl\ structure constants:
\eqn\eim{\eqalign{c_{t-1} &= x(2)c_t - \half g^{2,t+1}_3(0,0)c_{t+1}\cr
c_{t-1} &= -\half g^{2,t}_2(-1,0)c_t-\half g^{2,t+1}_3(-1,0)c_{t+1}~.\cr}}
The structure constants are simple and are provided in Appendix B. Plugging into \eim, taking their difference and iterating gives the following result for $x(t)$:
\eqn\eita{\eqalign{x(t) = \prod_{j=1}^{t-1}\left({j\over 2j-1}\right)\left(x(2)+{1-j\over 2}\right)~.\cr}}
To extract $x(2)$, we take a linear combination of the equations \eim\ that eliminates $c_{t+1}$ and set $t=2$. This yields the following quadratic equation for $x(2)$,
\eqn\eiub{x(2)^2+x(2) - {\l^2-1\over 4}=0}
with solutions $x(2) = -{1\pm\l\over 2}$, namely the expected Virasoro mode eigenvalues (up to a sign, cf. \sca). Plugging back into \eita\ gives the zero mode coefficients, $x(t)= 2\pi A_{\pm}(t)$, showing a match with CFT.

Finally, we obtain the coefficients $c_{t}$ by plugging $x(t)= 2\pi A_{\pm}(t)$ into \eij\ and substituting for the trace, giving
\eqn\eiug{c_{t} = {4^{t-1}(2t-1)\over (t-1)!}{\Gamma(\half)\over\Gamma({3\over 2}-t)}{\Gamma(\pm\l+1-t)\over\Gamma(\pm\l)}~.}
Therefore, the $A=0$ gauge propagator can be written as a series expansion \eil\ with expansion coefficients $c_{t}$ given in \eiug. This matches the result of \ChangMZ\ obtained by solving the equation \aa\ in AdS.

At $\l=1/2$ one can easily confirm our simple result \am\ using \aed, as well as the highest weight equations \eik\ using the formulae in Appendix A. Note also that acting on $c_{\pm}$ as\foot{These star products actually evaluate to give simple expressions in terms of Laguerre polynomials:
\eqn\dscb{y_1^{2m} * c_{\pm} * y_2^{2m} = (2m+n_{\pm})!(2i)^{2m}L_{2m+n_{\pm}}(2iy_1y_2)e^{-iy_1y_2}}
recalling that $n_-=0, n_+=1$.}
\eqn\dsca{y_1^{2m} * c_{\pm} * y_2^{2m}~, ~~ m\in \IZ_+}
gives the master fields which, upon transformation to AdS, contain propagators of fields dual to the {\it descendants} of $\Oc_{\pm}$ with conformal weights $h=\half \pm {1\over 4}+m$; this is evident from \ao.
\vs
 
In sum, scalar propagators  in any gauge field background obeying \aha\ can be obtained by gauge transformation of the same $A=0$ gauge master fields $c_{\pm}$, themselves highest weight states of \hsl.
In pure AdS these propagators obey the boundary conditions \ag.
Furthermore, there is a direct mapping between the entire highest weight representation of the CFT and bulk master fields in $A=0$ gauge, which we can demonstrate explicitly at $\l=1/2$; this is controlled by \hsl\ symmetry.

\subsec{Scalar propagators in general backgrounds}

We want a rule for obtaining the scalar propagator in an arbitrary higher spin background written in the gauge \aha.  Our working assumption is that the master fields $C_\pm$ are obtained by starting from $c_\pm$ in $A=0$ gauge, and then applying a gauge transformation.   The nontrivial claim here is that we can use the {\it same} $c_\pm$ for all backgrounds.

For asymptotically AdS backgrounds this claim is easy to justify.  While any $c_\pm$ will lead to a scalar propagator obeying the bulk wave equation, what distinguishes the specific $c_\pm$ under consideration is that it picks out the particular solution that obeys delta function boundary conditions.   But two asymptotically AdS backgrounds will differ by a gauge transformation that vanishes at infinity, and so the delta function boundary condition is preserved.

The case of non-asymptotically AdS backgrounds is a bit more subtle, since it is not clear a priori what asymptotic boundary condition the propagator should obey; in particular, so that we obtain sensible boundary correlation functions.   It is symmetry considerations and the topological nature of the pure higher spin theory that point to the rule described above.

If the bulk is to yield correlators that agree with those of CFT, we should expect to see the same structure controlling the two sides. The scalar field is dual to a  primary of a CFT whose interactions with the higher spin currents are fixed by the \hsl\ wedge symmetry. Likewise the interactions of $C_{\pm}$ with the higher spin fields via the Vasiliev field equations are fixed by the same symmetry. It is reasonable to expect that symmetry fixes a unique $A=0$ gauge master field $c_{\pm}$ which respects certain boundary conditions upon gauge transformation. This amounts to saying that the higher spin symmetry at the root of the Vasiliev equations fully fixes the dynamics of the scalar field at linearized order; indeed, this proved to be the case in earlier computations of three-point correlation functions from Vasiliev gravity \AmmonUA.

The topological nature of the higher spin sector also provides motivation for our perspective.  Any two higher spin backgrounds are related (locally) by a gauge transformation.  It would be surprising if the corresponding  scalar master fields were not related by the same gauge transformation. This leads to the conclusion that there is a unique scalar master field in $A=0$ gauge.

This idea will pass various tests in the remainder of the paper, where we consistently compute propagators in black hole and higher spin backgrounds.
\newsec{Scalar propagators in black hole backgrounds}

In this section we consider black hole backgrounds and the scalar propagators defined therein.  We  consider the case of the BTZ black hole, deferring the higher spin generalization to the next section.  The boundary two-point function obtained from the BTZ scalar propagator obeys the properties of a thermal CFT correlation function; in particular, it is periodic in imaginary time.   To set the stage for the generalization to the higher spin case, we show how this property is linked to the construction of black holes in higher spin language.

\subsec{BTZ black hole}

The BTZ connections are
\eqn\baa{\eqalign{a &= \left( V^2_1 +{1\over 4\tau^2} V^2_{-1}\right) dz \cr
\ab &= \left( V^2_{-1} +{1\over 4\taub^2} V^2_{1}\right) d\zb~.}}
The corresponding Euclidean BTZ metric is
\eqn\bab{   ds^2 =  d\rho^2 +{2\pi  \over k} \Big( \Lc dz^2 +\Lcb d\zb^2 \Big)+\Big( e^{2\rho}+\left({2\pi \over k}\right)^2\Lc \Lcb e^{-2\rho}\Big) dz d\zb~,}
where
\eqn\bac{\Lc = -{k\over 8\pi \tau^2}~,\quad \Lcb = -{k\over 8\pi \taub^2}~.}
$(\Lc,\Lcb)$ are the left and right moving components of the boundary stress tensor \BalasubramanianRE.    $\tau$ is the modular parameter of the Euclidean boundary torus, $z\cong z+2\pi \tau$.  To properly define a black hole we should also identify $z$ as $z\cong z+2\pi$; however, we will ignore this periodicity in our computations of scalar propagators, since it can be restored at the end by summing over images.

The black hole horizon is located at $\rho=\rho_+$, where
\eqn\baca{ e^{-2 \rho_+ }=4\tau\taub~.}

The scalar bulk-boundary propagator is
\eqn\bad{ G_\pm(\rho,{\bf x};{\bf 0}) =\pm {\lambda \over \pi}  \left( e^{-\rho} \over  e^{-2\rho} \cos \left(z\over 2\tau\right) \cos \left(\zb\over 2\taub\right) +4\tau\taub \sin \left(z\over 2\tau\right) \sin \left(\zb\over 2\taub\right) \right)^{1\pm \lambda}~.}

The propagator is invariant under the thermal identification $z\rt z+2\pi \tau$.  The propagator is also invariant under
\eqn\bae{  \rho \rt 2\rho_+ -\rho~,\quad z \rt z + \pi \tau~,}
which corresponds to a half  shift around the thermal cycle combined with reflection about the horizon.

These properties have important implications for the structure of boundary correlation functions computed from the Lorentzian BTZ solution.  The Lorentzian solution has an extended Kruskal diagram with two asymptotically AdS$_3$ regions, at $\rho \rt \pm \infty$.  In the AdS/CFT correspondence the Lorentzian BTZ solution is dual to an entangled state in a tensor product of two CFTs \refs{\BalasubramanianDE,\MaldacenaKR}; e.g.
\eqn\baf{ |\Psi\rangle = {\cal N} \sum_E e^{-{1\over 2} \beta E } |E\rangle_L \otimes |E\rangle_R}
in the static case, where $\beta = 1/T$ is the inverse temperature.  In computing the two-point function of scalar operators we have the option of taking the two operators to be associated with the same CFT factor, or with separate ones.   If both operators live in ${\cal H}_R$ associated with the boundary at $\rho \rt \infty$, then the two-point function is extracted from the large $\rho$ behavior of the propagator,
\eqn\bag{ \langle \Psi | {\cal O}_{\pm,R}(z,\zb) {\cal O}_{\pm,R}(0,0)|\Psi \rangle \sim  \left({1\over \sin \left(z\over 2\tau\right) \sin \left(\zb\over 2\taub\right) } \right)^{1\pm \lambda}~. }
On the other hand, suppose we replace $ {\cal O}_{\pm,R}(z,\zb)$ by  ${\cal O}_{\pm,L}(z,\zb)$.  The corresponding mixed boundary correlator is obtained from the $\rho \rt -\infty$ behavior of the propagator,
\eqn\bah{ \langle \Psi | {\cal O}_{\pm,L}(z,\zb) {\cal O}_{\pm,R}(0,0)|\Psi \rangle \sim  \left({1\over  \cos \left(z\over 2\tau\right) \cos \left(\zb\over 2\taub\right) } \right)^{1\pm \lambda} ~. }
As expected from the invariance of the propagator under \bae, the mixed correlator is equal to the single sided correlator \bag\ after a half shift, $z \rt z +\pi \tau$.

In Lorentzian signature we take $z=\phi+t$ and $\zb = \phi-t$, with $\phi$ and $t$ real. The single sided correlator \bag\ is singular on the light cone, $\phi = \pm t$.  These are the only singularities, since $\tau$ has a positive imaginary part.  By the same token, the mixed correlator \bah\ has no singularities on the Lorentzian section. This can be thought of as a consequence of the fact that the two asymptotic boundaries of the BTZ solution are causally disconnected.

\subsec{Scalar propagator on BTZ at $\lambda=1/2$}

We now wish to reproduce the result \bad\ using the techniques of higher spin gravity, so that we can generalize to bona fide higher spin black holes.   As explained in section 2, the propagator will be obtained as in \an, with the same, universal $A=0$ gauge master field $c_\pm$.    The details specific to BTZ are contained in the gauge transformation that generates the black hole.  In the BTZ case we have
\eqn\bba{\eqalign{\Lambda_\rho &= \left( e^\rho V^2_1 +{1\over 4\tau^2} e^{-\rho}V^2_{-1}\right) z \cr
\Lamb_\rho &= \left( e^\rho V^2_{-1} +{1\over 4\taub^2} e^{-\rho} V^2_{1}\right) \zb~,}}
and we recall
\eqn\bbb{\eqalign{\Phi_{\pm} =e^{(1\pm\l)\rho } \Tr \Big[e^{-\Lambda_\rho} \star c_\pm \star e^{\Lamb_\rho} \Big] ~. }}

Once again, we focus on the case $\lambda =1/2$  where the computation simplifies by using the Moyal product. Also, we restrict attention to the $\Phi_-$ case of alternate quantization; the standard quantization case is just slightly more messy. In terms of spinor variables we have
\eqn\bbc{\eqalign{ e^{-\Lambda_\rho}& =e_*^{{i\over 4} \left( e^\rho y_1^2 +{1\over 4\tau^2} e^{-\rho}y_2^2\right) z} \cr
 e^{\Lamb_\rho}& =e_*^{-{i\over 4} \left( e^\rho y_2^2 +{1\over 4\taub^2} e^{-\rho}y_1^2\right) \zb}~.  }}
On the right hand side, we have made it explicit that the functions are given by star-exponentials.   Our first task is to convert these to ordinary functions.   To achieve this we need to use the basic commutation relation $[y_1,y_2]_* =2i$ to put the star exponentials in Weyl ordered form.  Once this is done we use the fact that in a  Weyl ordered function the star product can be replaced by the ordinary product.

As explained in Appendix A.3, if $M$ is a symmetric $2\times 2$ matrix then
\eqn\bbd{ e_*^{{1\over 2} y^T M Y} = \sech(\sqrt{ |M|}) e^{{\tanh (\sqrt{|M|}) \over 2 \sqrt{|M|}} y^T M y}~. }
Using this result gives
\eqn\bbe{\eqalign{ \Phi_{-} =
& \quad e^{{1\over 2}\rho} \sech( Z) \sech( \Zb)\Tr \Big[e^{\tanh(Z) \left( \tau e^\rho y_1^2 +{1\over 4\tau} e^{-\rho}y_2^2\right) } * e^{-iy_1 y_2} *e^{\tanh (\Zb)  \left( \taub e^\rho y_2^2 +{1\over 4\taub} e^{-\rho}y_1^2\right)}  \Big]   }}
where we defined
\eqn\bbf{ Z = {iz\over 4\tau}~,\quad \Zb = {-i\zb \over 4\taub}~.}
To complete the computation we need the formula in Appendix A.2 for the star product of quadratic exponentials.  Using this, along with some elementary algebra, yields
\eqn\bbg{ \Phi_{-}  =  {e^{{1\over 2}\rho} \over \left( \cosh(2Z)\cosh (2\Zb) +4\tau \taub e^{2\rho} \sinh(2Z)\sinh(2\Zb) \right)^{1/2}}~. }
This is easily seen to lead to the propagator $G_-$ in \bad\ with $\lambda = 1/2$.

\subsec{Thermal periodicity and black hole holonomy}

In the Euclidean black hole geometry the thermal circle smoothly pinches off at the horizon.  This property is not immediately apparent when the solution is written in terms of the connections $A$ and $\Ab$, simply because the metric does not appear in this description.  Instead of considering the metric we can study the holonomy of the gauge connection around the thermal circle,
\eqn\bbh{ H= P e^{\oint A}~,\quad   \Hb= P e^{\oint \Ab}~.}
The analog of the statement that the thermal circle is smoothly contractible is that $H$ and $\Hb$ commute with all elements of the gauge algebra; that is, they lie in the center of the group we call \Hsl.  This is a key condition used in the construction of higher spin black holes: it is a gauge invariant condition, while the contractibility condition involving the metric is not.

Let us verify this in the case of the BTZ solution at arbitrary $\lambda$. We use the notation $(\cH,\overline{\cH})$ to denote the BTZ holonomy. Focussing on $\cH$, we have
\eqn\bbi{ \cH = e^{2\pi \tau \left(e^\rho V^2_1 +{1\over 4\tau^2}e^{-\rho}V^2_{-1} \right)}~.}
Now consider
\eqn\bbj{   \cH V^s_m \cH^{-1}~.}
The generators $V^s_m$ transform in the spin s-1 representation of the SL(2) subalgebra of \hsl, and so to compute \bbj\ we only need to use SL(2) representation theory.  The s=2 case can be evaluated using an explicit $2\times 2$ matrix representation for the SL(2) generators, and we readily compute
\eqn\bbj{   \cH V^2_m \cH^{-1}=V^2_m~.}
But since all other integer spin representations can be obtained by taking tensor products of the spin-1 representation, we immediately conclude that
\eqn\bbj{   \cH V^s_m \cH^{-1}=V^s_m}
for all $s$, which establishes that $\cH$ is in the center of \Hsl.   It is also straightforward to verify
\eqn\bbk{ \overline{\cH} = \cH~.}

It is now easy to establish that the scalar propagator -- indeed, the full scalar master field $C$ -- respects the thermal periodicity.  We note that
\eqn\bbl{ e^{-\Lambda_\rho}\Big|_{z+2\pi \tau} = e^{-\Lambda_\rho}\Big|_{z} \star \cH^{-1}~,\quad  e^{\Lamb_\rho}\Big|_{\zb+2\pi \taub} = e^{\Lamb_\rho}\Big|_{\zb} \star \cH~.}
Using this in \ap\ gives
\eqn\bbla{\eqalign{C_{\pm}(\rho,z+2\pi \tau,\zb+2\pi \taub) %&=e^{(n_\pm+{1\over 2})\rho }  \Big[\cH^{-1}\star e^{-\Lambda_\rho} \star c_\pm \star e^{\Lamb_\rho} \star \cH \Big] \cr
&= C_{\pm}(\rho,z,\zb)}}
as desired, and it follows trivially that
\eqn\bbm{ \Phi_{\pm}(\rho,z+2\pi \tau,\zb+2\pi \taub) =  \Phi_{\pm}(\rho,z,\zb)}

It is important that $C_{\pm}$, and not merely $\Phi_{\pm}$, is periodic: the higher components of $C$ are expressed as spatial derivatives of the trace $\Phi_{\pm}$, so \bbla\ ensures that all physical fields are completely smooth at the horizon.

\subsec{BTZ holonomy at $\lambda=1/2$}

It is instructive to obtain an expression for the holonomy $\cH$ in terms of spinor variables.   We have
\eqn\bbn{ \cH = e_*^{-{i\pi \over 4} (2\tau e^\rho y_1^2 +{1\over 2\tau}e^{-\rho} y_2^2)}~.}
We want to convert this into a function defined with ordinary products.  It's convenient to introduce
\eqn\bbn{ a ={1\over 2} \left( \sqrt{2\tau e^\rho} y_1 + {i\over \sqrt{2\tau e^\rho} } y_2\right)~,\quad a^\dagger = {1\over 2} \left( \sqrt{2\tau e^\rho} y_1 - {i\over \sqrt{2\tau e^\rho} } y_2\right)}
which obey
\eqn\bbo{ [a,a^\dagger]_* =1~. }
We then have
\eqn\bbp{ \cH = -ie_*^{-i\pi a^\dagger *a }~.}
Using the formulas in Appendix A.3 then gives
\eqn\bbq{ \cH = -2\pi i \delta^{(2)}(y)~.}

Direct computation using \aeb\ gives
\eqn\bbr{ \delta^{(2)}(y)*f(y) = f(-y) *  \delta^{(2)}(y)~.}
Since the generators of \hsl\ are all even functions of the spinor variables, \bbr\ shows that the spinor delta function  commutes with all elements of \hsl.

Another simple computation yields
\eqn\bbs{ \cH * \cH = -1~.}
Thus we have a $\IZ_4$ group of elements, $\{ 1, -1, \cH, -\cH\}$, that commute with all elements of hs[$\half$]: this defines a subgroup of the center of HS[$\half$]. We discuss this result further in section 5.

\subsec{Thermal periodicity for more general black holes}

We can now try to extend our discussion to more general black holes, in particular those carrying higher spin charges.    Let us first recall the rules for constructing black holes that are continuously connected to BTZ.   The first step is to write down flat connections that generalize BTZ to include additional higher spin chemical potentials and charges.   The holonomies around the thermal cycle can be written as
\eqn\bbs{ H = e^{\omega}~,\quad \Hb = e^{\omb}~.}
To fix the relations between the potentials and charges we demand that the holonomy is equivalent, up to conjugation by the gauge group, to that of BTZ.  This requirement can be implemented by demanding
\eqn\bbt{ \Tr (\omega^n) =  \Tr (\omega_{\rm BTZ}^n)~,\quad \Tr (\omb^n) =  \Tr (\omb_{\rm BTZ}^n)~, \quad n =1, 2, 3, \ldots~.}
The resulting black holes have sensible thermodynamic properties that precisely match those of a dual CFT description, when available.

Now, since $H$ is equal to $\cH$ up to conjugation, and since $\cH$ is a central element of the gauge group, it follows that $H= \cH$. And similarly  $\Hb= \cH$.

With this result in hand, we can easily prove that the scalar master field, hence the scalar propagator, computed in any such black hole will obey thermal periodicity. To show this we consider again the definition of $C$,
\eqn\bbba{\eqalign{C_{\pm}(\rho,z,\zb) =e^{(1\pm\l)\rho }  \Big[e^{-\Lambda_\rho} \star c_\pm \star e^{\Lamb_\rho} \Big]  }}
where

\eqn\bbu{\eqalign{ \Lambda_\rho &= A_z(\rho)z + A_{\zb}(\rho)\zb \cr \Lamb_\rho &= \Ab_z(\rho)z + \Ab_{\zb}(\rho)\zb~.}}
Flatness implies $[A_z, A_{\zb}]= [\Ab_z, \Ab_{\zb}]=0$. Using this, together with the definition of the holonomy, we have
\eqn\bbv{\eqalign{ C_{\pm}(\rho,z+2\pi \tau ,\zb+2\pi \taub)&  = e^{(1\pm\l)\rho }  \Big[H^{-1}\star e^{-\Lambda_\rho} \star c_\pm \star e^{\Lamb_\rho}\star H \Big] \cr
& = e^{(1\pm\l)\rho }  \Big[e^{-\Lambda_\rho} \star c_\pm \star e^{\Lamb_\rho} \Big]\cr
&= C_{\pm}(\rho,z,\zb)~,}}
which establishes periodicity, and provides an important consistency check on the interpretation of putative higher spin black hole solutions.

\newsec{Scalar propagator in the higher spin black hole background}

We will now derive the scalar propagator in the higher spin black hole background of \KrausDS, a brief summary of which is in order. This solution --- which we write below in subsection 4.2 --- has a spin-3 chemical potential, $\a$, which sources an infinite tower of higher spin charges, as well as their barred counterparts. These charges are fixed by the holonomy constraint \bbt, which has been solved perturbatively in $\a$ through $O(\a^{8})$, and there appears to be no obstruction to solving it to any order in $\a$.  The thermal partition function matches precisely with a dual CFT calculation in a generalized Cardy limit in which $\a\rar 0$ for fixed $\a/\t^2$ \refs{\KrausDS,\GaberdielYB}.

Until now, the charge assignments were only understood by matching trace invariants of $\o$ to those of $\omega_{\rm BTZ}$; but the analysis of the previous section tells us that if $\o=e^{-X}\star \o_{BTZ}\star e^X$, the  higher spin black hole holonomy will in the center of HS$[\lambda]$. This is indeed the case, and we show the details of this short calculation in appendix C.1 through $O(\a^2)$.

Before tackling the black hole, it will be helpful to first acquaint ourselves with higher spin deformations of the AdS and BTZ solutions using a simpler example in which we can compute the propagator nonperturbatively in a higher spin deformation.

\subsec{Higher spin warmup: Chiral spin-3 deformation}
The following chiral spin-3 perturbation of the AdS connection, previously studied in \AmmonUA, is a solution of the \hsl\ theory for finite $\mu$:
\eqn\gnc{\eqalign{a &= V^2_{1}dz-\mu V^3_{2}d\zbar\cr
\ab &= V^2_{-1}d\zbar \cr}}
where $\mu$ is a constant. This is the bulk dual to the finite deformation of the CFT action by a dimension-(3,0) operator, $\mw(z)$, with constant coupling: $\delta S_{CFT} = \mu \int d^2z \mw(z)$. It is useful to bear in mind that the higher spin black hole with chiral charge can be thought of as a finite temperature version of this solution.

In \AmmonUA, it was shown that the linearized scalar obeys the following equation of motion:
\eqn\gnd{\left[\p_{\rho}^2 + 2\p_{\rho}+4e^{-2\rho}(\p\overline{\p}-\mu \p^3) -(\l^2-1)\right]\Phi=0~.}
We use the methods of the previous sections to derive its bulk-to-boundary propagator at $\l=1/2$. We have, in oscillator notation,
\eqn\gne{\eqalign{\Lr &= {z\ep\over 4i}y_1^2+{\mu\zb e^{2\rho}\over 16}y_1^4\cr
\Lbr &= {\zb\ep\over 4i}y_2^2}}
and thus
\eqn\gnf{\Phi_{\pm} = e^{(n_{\pm}+\half)\rho}\Tr[e^{-{z\ep\over 4i}y_1^2-{\mu\zb e^{2\rho}\over 16}y_1^4}* c_{\pm}* e^{{\zb\ep\over 4i}y_2^2}]~.}
Note that each exponential depends only on a single oscillator, hence there is no distinction between the ordinary and star exponentials. For simplicity, we focus on $\Phi_-$ and use the integral representation of the Moyal product to evaluate the trace, which can be boiled down to the following integral:
\eqn\gng{\eqalign{\Phi_- &= {e^{\rho/2}\over 2\pi}\int du_1 e^{{iz\ep}u_1^2-{\mu\zb e^{2\rho}}u_1^4}\int dv_2e^{-{i\zb\ep\over 4}v_2^2+iu_1v_2}~.\cr}}
The $v_2$ integral is Gaussian, and the remaining integral provides a representation of a modified Bessel function,
\eqn\gnh{\int^{\infty}_{-\infty} dx ~e^{ax^2+bx^4}= \half \sqrt{a\over b}e^{-a^2\over 8b}K_{{1\over4}}\left({-a^2\over 8b}\right)~.}

Plugging in gives the final result written as a multiplicative correction to the AdS propagator,
\eqn\gni{\Phi_- = \Phi_-^{AdS} \times\sqrt{2y\over \pi }e^{y}K_{{1\over4}}\left(y\right)}
where
\eqn\gneab{y = -{(1+z\zb e^{2\rho})^2\over 8\mu \zb^3 e^{4\rho}}~.}
This indeed obeys \gnd.   The extra factor appearing in \gni\ modifies the asymptotic behavior of the propagator by a $(z,\zb)$ dependent factor.   We take this asymptotic behavior to define what we mean by the scalar bulk-boundary propagator in the presence of the spin-3 field.  From the asymptotic behavior we read off the boundary two-point function (not keeping track of the normalization)
\eqn\gneac{ \langle \Oc_-(z,\zb) \Oc_-(0,0)\rangle \sim \langle \Oc_-(z,\zb) \Oc_-(0,0)\rangle_{\mu=0} \times \sqrt{2y_\infty \over \pi }e^{y_\infty }K_{{1\over4}}\left(y_\infty\right) }
with
\eqn\gnead{ y_\infty = -{z^2 \over 8\mu \zb}~.  }

A similar result holds for $\Phi_+$, and both quantizations can be usefully expressed as series expansions in $\mu$,
\eqn\gnka{\Phi_{\pm} = \left[e^{\rho}\left({1\over 1+z\zb e^{2\rho}}\right)\right]^{1\pm\half}\sum_{n=0}^{\infty}c_{n,\pm}\left[\mu \zb^3e^{4\rho}\left({1\over 1+z\zb e^{2\rho}}\right)^{2}\right]^n}
with the coefficients
\eqn\gnf{c_{n,\pm} = {(4n\pm1)!!\over 4^n n!}~.}

\subsec{Higher spin black hole}
The unbarred connection for generic $\l$ is\foot{Compared to \KrausDS, a normalization factor $N(\l)$ appearing in $(a,\ab)$ has been removed. This merely rescales charges at $O(\a^2)$ and  beyond; to restore the conventions of \KrausDS, simply rescale $(\mw,\mwb)\rar N(\l)\cdot (\mw,\mwb)$ and $(\a,\alphab)\rar N(\l)\cdot(\a,\alphab)$ in the above.}
\eqn\pge{\eqalign{a_z &= V^2_{1} -{2\pi\ml\over k}V^2_{-1} -{\pi\mw\over 2k}V^3_{-2} + J\cr
a_{\zb}&= -{\mu} \left(a_z\star a_z -~{\rm trace}\right) ~.\cr}}
$\mw$ is the spin-3 charge. The object
\eqn\pgf{J = J_4V^4_{-3}+J_5V^5_{-4}+\ldots}
allows for an infinite series of spin-$s$ charges $\lbrace J_s\rbrace $, and $\mu$ is the spin-3 chemical potential. The solution is accompanied by the analogous  barred connection,
\eqn\pgeb{\eqalign{\ab_{\zb}&= V^2_{-1} -{2\pi\mlb\over k}V^2_{1} -{\pi\mwb\over 2k}V^3_{2} + \overline{J}\cr
\ab_z &= -{\mub} \left(\ab_{\zb}\star \ab_{\zb} -~{\rm trace} \right)  \cr}}
with the barred analog of \pgf. We will need both barred and unbarred sectors, but will often refer to only one for convenience.

To any order in $\mu$, one can recover the chiral deformation connection \gnc\ by setting $\mub=0$ and taking a limit in which $\tau\taub\rar\infty$ for fixed $\mu$; we will use this as a check on our work.

The chemical potentials that enter into the partition function, conjugate to the spin-3 charges $(\mw,\mwb)$, are labeled $(\a,\alphab)$ respectively and are related to $\mu$ by
\eqn\pgea{\a=\mu\taub~, ~~ \alphab = \mub\t~.}
The charges through $O(\a^8)$ are given in \KrausDS; we will work only at $O(\a)$, where the charges are given by
\eqn\cha{\eqalign{\ml &= -{k\over 8\pi\t^2} + O(\a^2)\cr
\mw &= -{k\over 3\pi\t^5}\a+O(\a^3)\cr
J_s &= O(\a^{s-2})}}
and analogously for the barred charges.

The scalar propagator $\Phi$ is given as a perturbation sum around the BTZ result, $\Phi^{(0)}$:
\eqn\pgb{\eqalign{\Phi_{\pm} &=e^{(n_{\pm}+\half)\rho}\Tr\left[e^{- \Lr}\star c_{\pm}\star e^{\Lbr}\right]\cr
&=\Phi_{\pm}^{(0)}+ \sum_{n=1}^{\infty}\Phi_{\pm}^{(n)} }}
where
\eqn\pga{\eqalign{\Lr &= \Lr^{(0)} + \sum_{n=1}\a^n \Lr^{(n)}\cr
\Lbr&= \Lbr^{(0)} + \sum_{n=1}\alphab^n \Lbr^{(n)}~.}}
Our goal is to compute the first order correction, $\Phi^{(1)}_{\pm}$. To isolate this piece we Taylor expand about $\a=0$ and rely on the following formula:
\eqn\pgc{{d e^{\Lr}\over d\a} = \int^1_0 ds ~e^{s\Lr}\star {d \Lr\over d\a}\star  e^{(1-s)\Lr}~,}
so that
\eqn\pgd{\eqalign{\Phi^{(1)}_{\pm} =e^{(n_{\pm}+\half)\rho}\int^1_0 ds&~\Tr\Big[\a \left(e^{- s\Lr^{(0)}}\star (-\Lr^{(1)})\star   e^{- (1-s)\Lr^{(0)}}\star c_{\pm}\star e^{\Lbr^{(0)}}\right) \cr&+ \overline{\a}\left(e^{- \Lr^{(0)}}\star c_{\pm}\star e^{s\Lbr^{(0)}}\star \Lbr^{(1)}\star e^{(1-s)\Lbr^{(0)}}\right) \Big]~.}}

We specialize to $\l=1/2$ for the remainder of this section and compute $\Phi_-$. Translating to oscillators, the connection is simple at $O(\a)$:
\eqn\pgh{\eqalign{a_z &= {1\over 4i}y_1^2 +{1\over 16i\t^2}y_2^2 -{\a\over 96\t^5}y_2^4\cr
a_{\zb}
&= {\a\over\taub}\left({1\over 16}y_1^4+{1\over 256\t^4}y_2^4+{1\over 32\t^2}(y_1y_2)^2\right)\cr
\ab_{\zb} &= {1\over 4i}y_2^2 +{1\over 16i\taub^2}y_1^2 -{\overline{\a}\over 96\taub^5}y_1^4\cr
\ab_{z}
&= {\overline{\a}\over\t}\left({1\over 16}y_2^4+{1\over 256\taub^4}y_1^4+{1\over 32\taub^2}(y_1y_2)^2\right)~.\cr}}
The linearized corrections to $(\Lr,\Lbr)$ are given by
\eqn\pgk{\eqalign{\Lr^{(1)} &= {\zb\ept\over 16\taub} y_1^4+{\zb\over 32\t^2\taub}(y_1y_2)^2+\left({\zb\over 256 \t^4\taub} - {z \over 96 \t^5}\right)e^{-2\rho}y_2^4\cr
\Lbr^{(1)} &= {z\ept\over 16\t} y_2^4+{z\over 32\taub^2\t}(y_1y_2)^2+\left({z \over 256 \taub^4\t} - {\zb \over 96 \taub^5}\right)e^{-2\rho}y_1^4\cr}}
and $(\Lr^{(0)},\Lbr^{(0)})$ were defined in \bba.

The easiest strategy for evaluating \pgd\ is to combine the exponentials --- all of which can be converted into ordinary functions --- into a single exponential. This final exponential will be that of a quadratic form once more, and we trace this against $(\Lr^{(1)},\Lbr^{(1)})$ by expanding the exponent and picking out the relevant quartic pieces (equivalently, by performing 4d Gaussian integrals).  All of these steps are easily automated.

Introducing a temporary shorthand,
\eqn\pgna{\Lr^{(1)} = \b y_1^4 + \gamma(y_1y_2)^2+\delta y_2^4}
where
\eqn\pgmba{\beta = {\zb\ept\over 16\taub}~, ~~ \gamma = {\zb\over 32\t^2\taub}~, ~~\delta = e^{-2\rho}\left({\zb \over 256 \t^4\taub} - {z \over 96 \t^5}\right)~,}
we combine the exponentials and, including the relevant determinant factor, the integral becomes
\eqn\pgn{\eqalign{\Phi_- &=  -\a e^{\rho/2}\sech(\Zb)\int^1_0 ds~\sech(sZ)~\sech((1-s)Z)\cr&\times~{1\over \sqrt{|L|}}\Tr\Big[(\b y_1^4 + \gamma(y_1y_2)^4+\delta y_2^4)\star e^{\half y^TSy}\Big]}}
where $(Z,\Zb)$ were defined in \bbf. Both $S$ and $|L|$ depend non-trivially on the parameter $s$ and are straightforward linear combinations of trigonometric functions, though we refrain from showing the full expressions here. Just to give the reader a taste, they simplify considerably in the $\rho\rar+\infty$ limit, for instance:
\eqn\pgr{\eqalign{S_{11}|_{\rho\rar+\infty} &\approx {4i\t \ep \sin\left({(1-s)z\over 2\t}\right)\sin\left({sz\over 2\t}\right)\over \sin\left({z\over 2\t}\right)}\cr
S_{22}|_{\rho\rar+\infty} &\approx -{i\left(\cos\left({z\over 2\t}\right)+\cos\left((s-\half){z\over \t}\right)\right)\over 2\t\ep\sin\left({z\over 2\t}\right)}\cr
S_{12}|_{\rho\rar+\infty}  &\approx -{i\sin\left({z(s-\half)\over \t}\right)\over \sin\left({z\over 2\t}\right)}\cr
|L||_{\rho\rar+\infty} &\approx {8\t\taub  e^{2\rho}\tan\left({\zb\over 4\taub}\right)\sin\left({z\over 2\t}\right)\over\cos^2\left({sz\over 4\t}\right)\cos^2\left({(1-s)z\over 4\t}\right)}~.
}}

Continuing with the calculation at arbitrary $\rho$, and isolating nonzero traces using
\eqn\pgua{\eqalign{\Tr[y_1^4\star e^{\half y^TSy}] &=  3S_{22}^2\cr
\Tr[(y_1y_2)^2\star e^{\half y^TSy}] &=S_{11}S_{22}+2S_{12}^2\cr
\Tr[y_2^4\star e^{\half y^TSy}] &= 3S_{11}^2\cr \cr}}
one arrives at the final result for the unbarred $O(\a)$ piece of the alternate quantization scalar propagator in the black hole background:

\eqn\pgce{\eqalign{\Phi_-^{(1)} = {i\a e^{\rho/2}\over 16\t^2}&\Big[\cosh^2(2\Zb) \left(-4(Z+\Zb)(\cosh(4 Z)-2)-\sinh(4Z)\right)\cr&+4e^{2\rho}\t\taub \sinh(4\Zb)\left(-4(Z+\Zb)\sinh (4Z)+2(\cosh (4Z)-1)\right)\cr&-(4e^{2\rho}\t\taub)^2\sinh^2(2\Zb)\left(4(Z+\Zb)(\cosh(4 Z)+2)-3\sinh (4Z)\right)\Big]\cr&\times \left(\cosh(2Z)\cosh(2\Zb)+4e^{2\rho}\t\taub\sinh(2Z)\sinh(2\Zb)\right)^{-5/2}~, }}
where $(Z,\Zb)$ were defined in \bbf.
This is manifestly periodic under thermal identification of $(z,\zb)$ in accord with the previous section's results.

As an immediate check, we consider the chiral deformation limit, keeping $\alphab=0$ and taking $\t\taub\rar \infty$. This sets all charges to zero, and the leading term should match  the $O(\mu)$ piece of \gnka. Indeed, in this limit
\eqn\pgcf{\Phi_-^{(1)} \approx \Phi_-^{BTZ}\times \left[\left({3\mu\zb^3e^{4\rho}\over 4}\right)\left({1\over 1+e^{2\rho}z\zb}\right)^2+\ldots\right].}

Following an identical procedure, one obtains the barred, $O(\alphab)$ piece of the propagator:
\eqn\pfa{\eqalign{\Phi_-^{(1)} = {i\overline{\a} e^{\rho/2}\over 16\taub^2}&\Big[\cosh^2(2Z)\left(-4(Z+\Zb)(\cosh(4 \Zb)-2)-\sinh (4\Zb)\right)\cr&+4e^{2\rho}\t\taub \sinh(4 Z)\left(-4(Z+\Zb)\sinh(4 \Zb)+2(\cosh (4\Zb)-1)\right)\cr&-(4e^{2\rho}\t\taub)^2\sinh^2(2Z)\left(4(Z+\Zb)(\cosh( 4\Zb)+2)-3\sinh (4\Zb)\right)\Big]\cr&\times \left(\cosh(2Z)\cosh(2\Zb)+4e^{2\rho}\t\taub\sinh(2Z)\sinh(2\Zb)\right)^{-5/2} ~.}}

The full propagator for nonzero $(\a,\alphab)$ is the sum of \pgce\ and \pfa. We see that the barred piece can be obtained from the unbarred piece by performing the operations
\eqn\pfc{Z\leftrightarrow \Zb~, ~~ \a \leftrightarrow \alphab}
which is equivalent to taking
\eqn\pfc{ z \leftrightarrow \zb~, ~~ \tau \leftrightarrow -\taub~, ~~ \mu\leftrightarrow
 -\mub~.}
Given that we define
\eqn\pfd{z = \phi+t~, ~~ \zb = \phi-t}
for real Lorentzian time $t$, this implies that, as expected, the propagator is invariant under time reversal in the static black hole background.

\subsec{Physical interpretation}
As we now elucidate, our results serve as further evidence that the higher spin black hole connection \pge\ and \pgeb\ should be fundamentally considered a black hole: the scalar field, whose coupling to the higher spin fields is fully determined by higher spin symmetry,  sees a thermal geometry with a well-behaved event horizon.\vs

\bul {\it Boundary CFT correlators and singularity structure}\vs

We read off the $O(\a)$ contribution to the black hole two-point boundary correlators $\langle \Oc(z,\zb)\Oc(0)\rangle$, by taking large $|\rho|$ limits. At $\rho\rar+\infty$, we have
\eqn\pgxa{\Phi_{-}^{(1)}|_{\rho\rar+\infty} \approx {\a e^{-\rho/2}\over 16\tau^2}{
3\sin{z\over \t}+(2+\cos{z\over \t})({\zb\over \taub}-{z\over \t})
\over\sin^2{z\over 2\t}\sqrt{4\t\taub \sin{z\over 2\t}\sin{\zb\over 2\taub}}}~.}
This gives the correlator in which both operators are on the same boundary. There are no singularities away from the origin and its thermal images, and near the origin,
\eqn\phc{\Phi_{-}^{(1)} \approx \Phi_{-}^{BTZ}\times \left({3\mu\over 4}{\zb\over z^2}+\ldots\right)~.}
Evidently the irrelevant spin-3 deformation of the CFT strengthens the short-distance singularity.

At $\rho\rar-\infty$, we have
\eqn\pgxc{\Phi_{-}^{(1)}|_{\rho\rar-\infty} \approx {\a e^{\rho/2}\over 16\tau^2}{
\sin{z\over \t}+(2-\cos{z\over \t})({\zb\over \t}-{z\over \t})
\over\cos^2{z\over 2\t}\sqrt{\cos{z\over 2\t}\cos{\zb\over 2\t}}}~.}
giving the mixed boundary correlator.  This result has implications for understanding the causal structure of our solution.
Recall \AmmonNK\ that by calling our solution a ``black hole'', we mean that the connection \pgh\ can be transformed to a gauge in which the geometry manifestly possesses an event horizon. As it stands, the geometry built from \pgh\ is a traversable wormhole connecting two asymptotic regions and a thermal horizon appears absent. The benefit of this gauge is that one knows how to compute boundary correlation functions, by identifying terms in the connection with boundary charges and chemical potentials \GutperleKF.

Taking its metric at face value, the two boundaries of the wormhole geometry are causally connected.   Were we to add in a scalar field with minimal coupling to the metric, and no coupling to the higher spin field,  we might expect to see singularities in the mixed correlator associated with signal propagation from one boundary to another.   But of course higher spin symmetry is incompatible with a minimally coupled field, and instead dictates specific coupling to the higher spin fields.  Thus the physical causal structure has to take into account the full set of background fields, not just the metric.
Our main observation is that evaluated on the Lorentzian section $(z=\phi+t,~ \zb = \phi - t)$, the mixed boundary correlator \pgxc\ is nonsingular.  This is compatible with an interpretation in which the two boundaries are causally disconnected, generalizing the fact that the two boundaries of Lorentzian BTZ are spacelike separated.  We take this as further  evidence for the thermal nature of the black hole solution.\vs

\bul {\it Half shifted correlator}\vs

For the BTZ black hole, one can obtain the mixed correlator by reflecting the one-sided correlator about the horizon $e^{\rho_+} = {1\over \sqrt{4\t\taub}}$ and performing a shift of $(z,\zb)$ by half a period. Interestingly, this is {\it not} correct in the present case. Performing this operation on the result \pgxa\ nearly yields \pgxc, only with $\sin {z\over \t} \rar -3 \sin {z\over \t}$. There is no fundamental reason  why the half shift should reproduce the actual mixed correlator, but perhaps there is something to be learnt from this near-miss.\vs

\bul {\it Behavior at the horizon}\vs

For the Euclidean BTZ black hole, the scalar propagator becomes independent of Euclidean time at the horizon, consistent with the fact that the Euclidean time circle shrinks to zero size there. We wish to ask whether $\Phi_-$ in the higher spin background is also time-independent at the horizon; in doing so, we must be careful about gauge-dependence. In particular, $\Phi_-$ should only be expected to be time-independent at $\rho=\rho_+$ in the black hole gauge.

Considering the static case ($\overline{\a}=\a,  \taub=-\t = -i\t_2$), the wormhole gauge propagator $(\Phi)$ evaluated at the horizon is
\eqn\pfe{\Phi_-^{(1)}|_{\rho_+}= -{\mu\over 16\sqrt{2}\t_2^{5/2}(\cosh{\phi\over\t_2})^{5/2}}\left(8\phi-4\phi\cosh{2\phi\over\t_2}+3\t_2\sinh{2\phi\over\t_2}-8\t_2\cosh{t\over\t_2}\sinh{\phi\over\t_2}\right)}
which is time dependent.
Fortunately, at $O(\a)$ we know the explicit black hole gauge transformation, and in Appendix C.2, we show that the gauge variation of $\Phi_-$ precisely cancels the time-dependent piece of \pfe.
The scalar propagator is therefore smooth at the horizon.\vs

\bul {\it Relation to CFT}\vs

Via the duality \GaberdielPZ\ we can in principle compare our result for the black hole correlator against a CFT computation.  The CFT is taken to live on a torus with modular parameter $\tau$, and its action is perturbed by the operator $\delta S_{\rm CFT} = \mu \int\! d^2w~ \Wc(w)$, where $\Wc(w)$ is the spin-3 current.    Working to linear order in $\mu$, we should thus compare our first order correction against
\eqn\pra{  \mu \int\! d^2w \langle \Wc(w) \Ocb(z,\zb) \Oc(0,0)\rangle_{\rm CFT}~. }
We will not attempt the challenging task of computing this correlator, but instead just show that the leading short distance behavior found in \phc\ is compatible with that of \pra.

To keep the discussion as simple as possible, we consider the simplest theory with a spin-3 current, namely a complex free boson $\phi$. The spin-3 current is \BakasRY\
\eqn\prb{ \Wc = \p^2 \phib \p \phi - \p \phib \p^2 \phi~.}
As a scalar primary we take $\Oc = e^{ik\phi}$.   Consider the path integral expression for the scalar two-point function in the unperturbed CFT
\eqn\prc{\langle\Ocb(z,\zb)\Oc(0,0) \rangle= \int\! {\cal D}\phi {\cal D}\phib  e^{\int\! d^2 w \p \phi \overline{\p} \overline{\phi}}\Ocb(z,\zb)\Oc(0,0)~. }
We take the CFT to be defined on the plane, which should give the same short distance behavior as on the torus.

Now perform the change of variables, $\phib \rt \phib + \mu \zb \p^2 \overline{\phi}$, recalling that we treat $\phi$ and $\phib$ as independent variables in the path integral.  Expand \prc\ to first order in $\mu$.  After an integration by parts and using the zeroth order equations of motion  we find that the change in the action is
\eqn\hapd{ \delta \int\! d^2 w \p \phi \overline{\p} \overline{\phi} \sim \mu \int\! d^2w \Wc(w)~. }
The change in the scalar operator is
\eqn\hape{ \delta \Ocb(z,\zb) \sim \mu k \zb \p^2  \overline{\phi}(z,\zb) \Ocb(z,\zb)~.}
Since this was just a change of variables the total change is zero, so
\eqn\hapf{ \langle  \int\! d^2 w \Wc(w) \Ocb(z,\zb)\Oc(0,0) \rangle \sim  k \zb  \langle  \p^2  \overline{\phi}(z,\zb) \Ocb(z,\zb)\Oc(0,0)\rangle \sim k^2{\zb \over z^2} \langle\Ocb(z,\zb)\Oc(0,0) \rangle  }
where the last step just comes from applying Wick's theorem.  Comparing the appearance of the prefactor $\zb/z^2$ to that in \phc, we see that we recover the same structure.

\newsec{Comments on conical defects and the center of \Hsl}
In the preceding sections we established that the BTZ (and higher spin) black hole thermal holonomy $\cH$ --- which is proportional, at $\l=1/2$, to a two-dimensional spinorial delta-function $\delta^{(2)}(y)$ --- commutes with all generators of \hsl. Our result, as a mathematical statement independent of black holes, is that $\cH$ is an element of the center of \Hsl, the group exponentiation of \hsl.

Conical defect solutions of SL(N) higher spin gravity were studied in  \CastroIW, based on writing down flat connections whose holonomies around the angular direction lie in the center of SL(N).   We can aim to write down analogous solutions in the \Hsl\ theory.  Here we just make a few observations, leaving a more comprehensive study for the future.

\subsec{Simple conical defects in \hsl\ gravity}
The \Hsl\ group element $\cH$ was defined in \bbi\ as
\eqn\csd{\cH = e^{\pi(2\t\ep V^2_1 + {1\over 2\t\ep}V^2_{-1})}~.}
$\cH$ lies in the center of \Hsl\ for any $2\t\ep$; indeed $\cH$ is independent of $2\t\ep$.  Since at the moment we are not considering black holes, redefine $2\t\ep=\gamma$ to be some arbitrary complex constant. In addition, it is obvious that integer powers of $\cH$ will also be central. Then we have the statement that the object
\eqn\csf{\cH^n = e^{n\pi(\gamma V^2_1 + {1\over \gamma}V^2_{-1})}~, ~~ n \in \IZ~, ~~ \gamma \in \Bbb{C}}
is a central element of \Hsl.  As above, there is actually no dependence on $\gamma$, but it will be convenient to include this fictitious parameter.   Smooth conical defect solutions are then obtained by writing down flat connections whose holonomy around the angular direction is $\cH^n$.

For instance, we can take
\eqn\csk{ a_{\phi}  = {n\over 2} \left(\gamma V^2_1 + {1\over \gamma}V^2_{-1}\right)~, ~~ n \in \IZ}
with $\phi \cong \phi +2\pi$, so that $e^{\oint a_\phi d\phi}= \cH^n$. In this language, global AdS corresponds to $n=1,\gamma=2$.\foot{Note that, at least at $\l=1/2$, the angular holonomy of global AdS is not simply the identity. This is somewhat analogous to the situation in SL(N,$\Bbb{C}$) gravity for $N$ even, where the global AdS holonomy is minus the identity.}

As discussed in \CastroIW, conical surplus solutions are more sensible in a Euclidean theory with a complex connection. Thus, take $\gamma$ in \csk\ to be be complex, and work in Euclidean time, with $z=\phi +it_E, \zb=\phi-it_E$. We have
\eqn\csoa{\ab= -a^{\dagger}}
and \hsl\ generators satisfy the hermiticity property
\eqn\csob{(V^s_m)^{\dagger} = (-1)^mV^s_{-m}~.}
In the present case, the connection \csk\ together with its barred partner is
\eqn\cso{\eqalign{a_{z} &= {n\over 2} \left(\gamma V^2_1 + {1\over \gamma}V^2_{-1}\right)\cr
\ab_{\zb} &= {n\over 2} \left(\gamma^* V^2_{-1} + {1\over \gamma^*}V^2_{1}\right)~.\cr}}
We can set $\gamma=1$ by shifting $\rho$ after conjugating the connections by $b$. The metric, using a convenient trace normalization, reads
\eqn\csp{ds^2 = d\rho^2 + {n^2\over 4}\left[(\ep+\emp)^2dt_E^2+(\ep-\emp)^2d\phi^2\right]~.}
The $\phi$ cycle closes off at $\rho_0 = 0$ where there is a surplus angle
\eqn\csr{\delta = 2\pi(|n|-1)~.}
Because we obtain this connection from that of BTZ by rescaling/relabeling, we can read off various physical quantities in this conical surplus background, for instance the scalar bulk-boundary propagator. As discussed in \CastroIW, these conical surplus solutions have an energy that lies below that of global AdS, and so are presumably not present in a consistent theory whose energy is bounded from below.

These solutions are rather trivial, in that the connection is just built out of SL(2) generators.  They are an example of the fact that any solution of ordinary (i.e. SL(2)) gravity can be lifted to a solution of \hsl\ gravity.  Much more interesting would be smooth solutions of the \hsl\ theory with no SL(2) counterpart.  Finding such solutions reduces to the following problem.  To be asymptotically AdS, these solutions should obey the Drinfeld-Sokolov boundary conditions,
\eqn\hwa{{a}_z = V^2_1 + \sum_{s=2}^{\infty}Q_s V^{s}_{1-s}~.}
We then ask: for what values of spin-$s$ charges $\lbrace Q_s\rbrace$ is the object $H = e^{2\pi {a}_z}$ in the center of \Hsl?

\subsec{Center of HS[$\half$]}

Elements of HS[$\half$] can be expressed as $e_*^{p(y_1,y_2)}$, where
$p(y_1,y_2)$ is an element of hs$[{1\over 2}]$.  In particular, $p(y_1,y_2)$ has a series expansion in nonnegative powers of $y_{1,2}$, obeying $p(-y_1, -y_2)=p(y_1,y_2)$ and $p(0,0)=0$.  Here we wish to find the center of HS[$\half$].

We first answer the closely related question of what is the most general even function $g(y_1,y_2)$ that star-commutes with all other even functions.  We use the shorthand $y=(y_1,y_2)$ and $yz = y_1 z_2 - y_2 z_1$.    A straightforward computation using the integral form of the star product yields

\eqn\cgc{[f(y),g(y)]_{*} = {i\over 2\pi^2}\int d^2u f(y+u)\int d^2v \sin(uv)g(y+v)~.}
After shifting variables, $v\rar v-y~, ~ u\rar u-y$ we obtain
\eqn\cgi{\eqalign{[f(y),g(y)]_{*} = 2i\int d^2u f(u)&\sin(yu)\int d^2v g(v)\sin(vy)\sin(vu)~.}}
For this to vanish for any even $f(u)$ we clearly need
\eqn\cgia{ \sin(yu)\int d^2v~ g(v)\sin(vy)\sin(vu)=0~}
for all $y$ and $u$.  Fourier transforming,
\eqn\cgk{ g(v) = \int\! {d^2k \over (2\pi)^2} \tg(k)\cos(vk)}
the condition \cgia\ becomes
\eqn\cgm{ \sin(yu) \left[\tg(y-u)-\tg(y+u)  \right]=0~. }
The general solution for $\gt(k)$ that satisfies \cgm\ for all all $y$ and $u$ is
\eqn\cgn{ \gt(k) =  \alpha + \beta \dtwo(k)~,}
where $\alpha$ and $\beta$ are constants.  The most general even $g(v)$ that star commutes with all even functions is therefore
\eqn\cgn{ g(y) = \alpha \dtwo(y) + \beta~. }

The remaining question is which such $g(y)$ can be written as $e_*^{p(y)}$.     We already know two examples,
\eqn\cgo{ g(y) = 1~, ~~\cH = -2\pi i \dtwo(y)~.}
Furthermore, since $\cH * \cH =-1$ (see \bbs), we also have $g(y) = -1,  -\cH$.  As discussed in subsection 3.5, these four elements generate  $\IZ_4$, and so we have established that the center of HS$[{1\over 2}]$ contains a $\IZ_4$ subgroup.  Whether the center is larger than this depends on whether there are additional functions of the form in \cgn\ that can be written as $e_*^{p(y)}$; preliminary investigations suggest an affirmative answer, but we reserve further comment at present.

\newsec{Conclusion}
Our study of scalar propagators in higher spin gravity lends significant support to the interpretation of the solution of \KrausDS\ as a bona fide black hole with higher spin charge. Let us mention a few immediate avenues for further research.

It would be nice to extend our explicit computations of scalar propagators to $\l\neq1/2$, which is an interesting problem physically and technically. No longer does \hsl\ have a Moyal product representation using oscillators $y_{1,2}$, but some analog of this is likely necessary: we know of no way to greatly simplify star exponentials of \hsl\ generators. One can write down ``deformed'' oscillators which satisfy a modified commutation relation and which one can use to construct \hsl\ generators \AmmonUA, but there is no clean way to take their star products. There has been some discussion of coherent states using deformed oscillators in the literature which one could hope to leverage in computing the simpler AdS and BTZ propagators \refs{\MukundaFV,\MehtaBY}\foot{The deformed oscillators are called ``para-Bose'' oscillators in these references.}. A successful generalization of our black hole computations to $\l\neq 1/2$ would allow full comparison to the dual CFT \GaberdielPZ, or to the free boson theory with $\mw_{\infty}[1]$ symmetry discussed in section 4.3.

Also on the CFT side, it would be desirable to reproduce the $O(\mu)$ black hole correlators of section 4. This would require a new CFT computation of integrated torus three-point functions between a spin-3 current and two scalar primaries.

The $A=0$ gauge method used here and in \AmmonUA\ has proven itself a powerful tool in understanding linearized scalar field dynamics in three dimensional Vasiliev theory. It can clearly be put to more use. Our goal herein was to compute bulk-boundary scalar propagators, but solutions of scalar field equations with other boundary conditions are clearly of interest and should be equally amenable to this treatment.

It is also be interesting to ask whether the \hsl\ theory, at non-integer values of $\l$, admits smooth classical solutions that have yet to be discovered. In particular, one might wish to search for smooth \hsl\ conical defect solutions. Before we can solve for the spectrum of smooth solutions and their higher spin charges, we must understand the center of \Hsl, itself a rich problem. Our results in section 5, in particular at $\l=1/2$, can already be utilized to unearth some (though probably not all) smooth conical defects of the hs[$\half$] theory. Such solutions would seem to have no immediate holographic relation to the light states of the $W_N$ minimal models in the 't Hooft limit -- indeed, being ``light'' in this limit, these states should not be dual to classical backgrounds in the bulk theory with $0\leq \l \leq 1$. Nevertheless, if smooth defect-like solutions exist they should have some dual CFT interpretation.

\vskip .3in

\noindent
{ \bf Acknowledgments}

\vskip .3cm

 This work was supported in part by NSF grant PHY-07-57702.
 We thank Tomas Prochazka and Joris Raeymaekers for discussions.

\appendix{A}{Useful identities involving Moyal products}

\subsec{Basic identities}
The following is a small collection of results involving spinors $y_{1,2}$. First, the following integral is useful for taking traces:
\eqn\czbc{{1\over 4\pi^2}\int d^2u d^2v e^{iuv}(v_1u_2)^{m}(v_2u_1)^{n} = m!~ n! ~(-i)^m i^n~.}
The following results are useful for evaluating star products against the object $c_-=e^{-iy_1y_2}$:
\eqn\cofc{\eqalign{f(y_1)\star e^{-i y_1y_2} &= f(2y_1)e^{-i y_1y_2}\cr
f(y_2)\star e^{-i y_1y_2} &= 0\cr
e^{-i y_1y_2}\star f(y_1) &= 0\cr
e^{-i y_1y_2}\star f(y_2) &= f(2y_2)e^{-i y_1y_2}~.\cr}}
Using
\eqn\acs{y_1^my_2^n = y_1^my_2^{n-1}\star y_2-imy_1^{m-1}y_2^{n-1}}
and the first two equations of \cofc, one can show the general result
\eqn\cofd{y_1^my_2^n\star \e=\left\{\matrix{  (-i)^n{m!\over (m-n)!}(2y_1)^{m-n}\e &   m \geq n \cr
  0 &   m < n } \right.  ~. }
\cofd\ is all one needs to show that $c_-$ at $\l=1/2$ is indeed a highest weight state of \hsl\ obeying equations \eik.

\subsec{Star product of quadratic exponentials}

We want to compute
\eqn\faa{ e^{{1\over 2}y^T M y} * e^{{1\over 2}y^T N y}}
where $M$ and $N$ are symmetric $2\times 2$ matrices.
From \aeb\ this is
\eqn\fab{ e^{{1\over 2}y^T M y} * e^{{1\over 2}y^T N y}= {1\over 4\pi^2}e^{{1\over 2}y^T (M+N) y}\int\! d^4u \exp \left\{{1\over 2}u^T L u  + J^T u \right\}  }
with
\eqn\fac{ L = \left( \matrix{M & -\sigma_2 \cr \sigma_2 & N} \right) }
and
\eqn\faca{  J^T = (y^T M, y^TN)~.}
Note that
\eqn\fad{L^{-1} = \left(\matrix{ \sigma_2 N X & -X^T \cr  -X & -\sigma_2 M X^T}\right)}
with
\eqn\fae{ X= (\sigma_2+ M\sigma_2 N)^{-1}}
and
\eqn\faf{|L|  = \det M \det N +\Tr(M \sigma_2 N \sigma_2)   +1~. }
Doing the integral in \fab\ gives
\eqn\fag{\eqalign{ e^{{1\over 2}y^T M y} * e^{{1\over 2}y^T N y} &= {1  \over \sqrt{|L|}}e^{{1\over 2}y^T (M+N) y-{1\over 2} J^T L^{-1} J  } \cr & ={1  \over \sqrt{|L|}}e^{{1\over 2}y^T S y}  }}
with
\eqn\fah{ S= \sigma_2 XM +MX^T N+NXM -\sigma_2 X^TN}

\subsec{Weyl ordering formulas}

Given some star function of $(y_1,y_2)$ we seek to Weyl order it so that it can be converted into an ordinary function.  A convenient method for Weyl ordering employs coherent states, as we now review.

Starting from $[y_1,y_2]_* =2i$ we define
\eqn\ha{ a= {y_1 + i y_2 \over 2}~,\quad a^\dagger = {y_1 -i y_2 \over 2}}
which obey $[a,a^\dagger]_*=1$.     A coherent state $|v\rangle$ obeys
\eqn\hb{ a |v\rangle = v |v\rangle~,}
and is normalized so that the completeness relation is
\eqn\hc{ 1 = {1\over \pi} \int\! d^2 v |v\rangle \langle v|~. }
Let $|n\rangle$ be the usual (unit normalized) number operator eigenstates.    We have
\eqn\hd{ \langle n |v\rangle = e^{-{1\over 2}|v|^2} {v^n \over \sqrt{n!}}}
so we can write
\eqn\he{ |v\rangle = e^{-{1\over 2}|v|^2} \sum_n {v^n\over \sqrt{n!}}|n\rangle=e^{-{1\over 2}|v|^2} e^{v a^\dagger }|n=0\rangle~. }
From this we easily derive
\eqn\hf{ \langle w |v\rangle = e^{-{1\over 2} |w|^2 -{1\over 2}|v|^2 + w^* v}}
and
\eqn\hg{ e^{-\lambda a^\dagger a}|v\rangle = e^{-{1\over 2} (1-|e^{-\lambda}|^2)|v|^2} | e^{-\lambda}v\rangle~.}
We also have
\eqn\hga{ \Tr (O) = {1\over \pi^2}\int\! d^2v d^2w e^{-{1\over 2}|v|^2-{1\over 2}|w|^2+w^* v}\langle v|O|w\rangle~.}

Now, let $G$ be some star function of $y_{1,2}$. Trading the spinors for raising/lowering operators, we can write $G=G(a,a^\dagger)$.  Weyl ordering consists of using the commutation relations to symmetrize the placement of $a$ and $a^\dagger$. Once this has been done, we can define a function of c-number variables, $G_W(v,v^*)$ via the replacement $a\rt v$, $a^\dagger \rt v^*$.   Finally, writing $v= (y_1+iy_2)/2$ gives us an ordinary function  that is equal to the original star function.

To implement this, we use that $G_W$ is given in terms of $G$ by  \AgarwalWC,
\eqn\ena{ G_W(v,v^*) = {2\over \pi} e^{2|v|^2} \int\! d^2 \alpha ~\langle -\alpha | G |\alpha\rangle e^{2(\alpha^* v - \alpha v^*)}}
where $|\alpha\rangle$ is a coherent state.

The following point will be important: to put an operator into Weyl ordered form we only use the commutation relation, and we don't need that $a^\dagger$ is the Hermitian conjugate of $a$.  So if we have two operators, $a$ and $b$,  obeying $[a,b]=1$, with $b$ not necessarily related to $a^\dagger$, then we can Weyl order this exactly as above; i.e we can pretend that $b=a^\dagger$.

A simple computation yields
\eqn\eea{G= e^{-\gamma(a^\dagger a + a a^\dagger)}\quad \Rightarrow \quad G_W(v,v^*)=    \sech( \gamma) e^{-2\tanh(\gamma)v^* v }~.}
In terms of our spinor variables and the star-exponential this reads
\eqn\eeab{ e_*^{-{\gamma \over 2}(y_1^2 + y_2^2)} =  \sech( \gamma) e^{-{1\over 2}\tanh(\gamma)(y_1^2 +y_2^2)}~.}

To derive a more general version we first note
\eqn\eeb{{1\over 2} \matrix{(y_1 & y_2) \cr &}*\left(\matrix{\cosh u &\sinh u \cr \sinh u &\cosh u}\right)\left(\matrix{y_1 \cr y_2 }\right) =( b b^\dagger + b^\dagger b)}
with
\eqn\eec{ \left( \matrix{ b \cr b^\dagger}\right) = \left(\matrix{ \cosh {u\over 2} & -i \sinh{u\over 2} \cr i\sinh{u\over 2}& \cosh{u\over 2} }\right)\left(\matrix{a \cr a^\dagger}\right)}
obeying
\eqn\eeca{[b,b^\dagger]=1~.}
Further, any symmetric matrix $M$ can be written as
\eqn\eed{ M = \sqrt{|M|} \left(\matrix{\cosh u &\sinh u \cr \sinh u &\cosh u}\right) }
and so
\eqn\eee{{1 \over 2} y^T M y =  \sqrt{|M|} ( b b^\dagger + b^\dagger b)~.}
Then from \eeab\ we derive the very useful result
\eqn\eef{  e_*^{{1\over 2} y^T M y } =  \sech ( \sqrt{|M|}) \exp \Big\{ {\tanh( \sqrt{|M|)} \over  \sqrt{|M|}}{1\over 2} y^T M y  \Big\}~.}

\appendix{B}{Some \hsl\ fundamentals}

The \hsl\ commutation relations among generators $\lbrace V^s_m\rbrace$ are
\eqn\vsa{[V^s_m,V^t_n] = \sum_{u=2,4,6,\ldots}^{s+t-|s-t|-1}g^{st}_u(m,n;\l)V^{s+t-u}_{m+n}}
with structure constants $g^{st}_u(m,n;\l)$ defined elsewhere, e.g. \refs{\PopeSR,\GaberdielWB, \KrausDS}. The commutation relations \vsa\ can be realized as
\eqn\bff{[V^s_m,V^t_n]= V^s_m \star V^t_n - V^t_n \star V^s_m}
if we define the associative product
\eqn\bg{V^s_m \star V^t_n \equiv \half \sum_{u=1,2,3,...}^{s+t-|s-t|-1}g^{st}_u(m,n;\l)V^{s+t-u}_{m+n}~.}
This is known as the ``lone star product'' \PopeSR.

The generators with $s=2$ form an SL(2,R) subalgebra, and the remaining generators transform simply under the adjoint SL(2,R) action as
\eqn\be{[V^2_m,V^t_n] = (m(t-1)-n)V^t_{m+n}~.}

We will need only the following four structure constants, and only in section 2:
\eqn\vsb{\eqalign{g^{2,t}_2(-1,0;\l) &= {1-t}\cr
g^{2,t+1}_3(0,0;\l)&={1\over 2}{(\l^2-t^2)t^2\over 4t^2-1}\cr
g^{2,t+1}_3(-1,0;\l)&= {1\over 2}{(\l^2-t^2)t(1-t)\over 4t^2-1}\cr
g^{t,t}_{2t-1}(0,0;\l) &= {4^{1-t}\Gamma(t)^2\over (2t-1)!!(2t-3)!!}
{\Gamma(\l+t)\over\l\Gamma(\l-t+1)}~.\cr}}
The last of these is proportional to the trace of two zero modes,
\eqn\vsc{\Tr(V^t_0V^t_0) = \half g^{t,t}_{2-1}(0,0;\l)~.}

\appendix{C}{Higher spin black hole calculations}
\subsec{Holonomy of higher spin black hole}

We show here that for the spin-3 black hole,
\eqn\cna{\o = e^{-X}\star \o_0\star e^X}
for $X \in$ \hsl, and hence that its holonomy $H=e^{\o}$ is in the center of the group \Hsl, $H=\cH$. Writing
\eqn\cnb{X = \a X_1 + \a^2X_2+\ldots}
and
\eqn\cnba{\o =\o_0 + \a\o_1+\a^2\o_2+\ldots}
we will confirm \cna\ through second order. In the normalization conventions of section 4 -- with the connection \pge, \pgf\ and potential \pgea\ -- the charges through $O(\a^2)$ are \KrausDS\
\eqn\ccbba{\eqalign{\ml &= -{k\over 8\pi\t^2} + {(\l^2-4)k\over24\pi\t^6}\a^2+O(\a^4) \cr
\mw &= -{k\over 3\pi\t^5}\a+O(\a^3)\cr
J_4 &= {7\over 36\t^8}\a^2+O(\a^4)~.}}
This yields
\eqn\cnbb{\eqalign{\o_0 &= 2\pi\t(V^2_1 + {1\over 4\t^2}V^2_{-1})\cr
\o_1 &= -2\pi (V^3_2 + {1\over 2\t^2}V^3_0-{5\over 48\t^4}V^3_{-2})\cr
\o_2 &= -2\pi (-{1\over 9\t^7}V^4_{-3}+{1\over 3\t^5}V^4_{-1} + {\l^2-4\over 60\t^5}V^2_{-1})~.\cr}}%where $N = \sqrt{20\over \l^2-4}$.

From \cna, we want to find $X$ such that
\eqn\cnca{\eqalign{\o_1 &= [\o_0,X_1]_{\star}\cr
\o_2 &= \half\lbrace \o_0,(X_1)_{\star}^2\rbrace_{\star}  - X_1\star \o_0\star X_1+ [\o_0,X_2]_{\star}~.}}
We find a solution with
\eqn\cnd{\eqalign{X_1 &= c^3_1 V^3_1 + c^3_{-1}V^3_{-1}\cr
X_2 &= c^4_2 V^4_2 + c^4_0 V^4_0 + c^4_{-2} V^4_{-2} + c^2_0 V^2_0\cr}}
where
\eqn\cne{\eqalign{c^3_1 =  -{1\over\t}~,& ~~ c^3_{-1} = - {5\over 12\t^3}~,\cr
c^4_2 = -{1\over \t^2}~, ~~ c^4_0 = -{2\over3\t^4}~,& ~~ c^4_{-2} = -{13\over 48\t^6}~, ~~ c^2_0 = {\l^2-4\over 15\t^4}~.}}

Extending to higher orders is a purely mechanical exercise, and it is only reasonable to expect \cna\ to hold to all orders.

\subsec{Black hole gauge at $O(\a)$}
Consider the higher spin black hole connection \pgh, written in wormhole gauge to $O(\a)$. Our goal in this subsection is to show that upon transforming to the black hole gauge, the scalar propagator evaluated at the horizon (cf. \pfe) is time-independent.

We will consider the static case, for which we know the gauge transformation explicitly. Due to the fact that the highest spin generator appearing in the connection is of spin-3, we can in fact directly inherit the transformation from the SL(3,$\Bbb{R}$) analysis of \AmmonNK.

Under an infinitesimal gauge transformation with \hsl-valued parameter $\chi = -\overline{\chi}$,
\eqn\pfg{\delta C = C\star \overline{\chi} - \chi\star C = -\lbrace\chi,C\rbrace_{\star}~.}
From \AmmonNK, we know to take
\eqn\pfh{\chi(\rho) = -{2\mu\over 3\t_2}\cosh(\rho-\rho_+)(V^3_1-V^3_{-1})}
where
\eqn\pfi{[V^2_i,V^3_m] = (2i-m)V^3_{i+m}~.}
The transformed connection will generate a manifestly smooth solution at $\rho=\rho_+$.
To linear order, we thus have the gauge transformation
\eqn\pfl{\delta C = {2\mu\over 3\t_2}\cosh(\rho-\rho_+)\lbrace V^3_1-V^3_{-1},C^{BTZ}\rbrace_{\star}}
and so the scalar propagator picks up a piece
\eqn\pfm{\delta\Phi_-^{(1)} ={4\mu\over 3\t_2}\cosh(\rho-\rho_+)\Tr\left[( V^3_1-V^3_{-1})\star C_-^{BTZ}\right]~.}
Even though $\chi(\rho)$ depends on $\rho$ alone, $\delta\Phi_-^{(1)}$ can pick up new dependence on $(\phi,t)$ via higher spinorial components of $C_-^{BTZ}$.

We are interested in the gauge transformation at the horizon. At $\l=1/2$, we can evaluate
\eqn\pfn{\eqalign{\Tr\left[( V^3_1-V^3_{-1})* C_-^{BTZ}|_{\rho_+}\right] &=
e^{\rho_+\over2}\sech\left({\phi+t\over 4\t_2}\right)\sech\left({\phi-t\over 4\t_2}\right)\Tr\left[( V^3_1-V^3_{-1})* {e^{\half y^TSy}\over\sqrt{|L|}}\right]\cr
&= {3\sech\left({\phi+t\over 4\t_2}\right)\sech\left({\phi-t\over 4\t_2}\right)\over 16\sqrt{2\t_2}\sqrt{|L|}}S_{12}(S_{22}-S_{11})}}
where
\eqn\pfo{\eqalign{S_{11} &= {i(\sinh{t\over\t_2}+\sinh{\phi\over\t_2})\over \cosh{\phi\over \t_2}}\cr
S_{22} &= {i(\sinh{t\over\t_2}-\sinh{\phi\over\t_2})\over \cosh{\phi\over \t_2}}\cr
S_{12} &= -{i\cosh{t\over\t_2}\over\cosh{\phi\over\t_2}}\cr
|L| &= {4\cosh{\phi\over\t_2}\over \left(\cosh{\phi\over2\t_2}+\cosh{t\over2\t_2}\right)^2}~.}}
Plugging into \pfm\ and using
\eqn\pfq{\cosh(2a)+\cosh(2b) = 2\cosh(a+b) \cosh(a-b)}
we get
\eqn\pfp{\delta\Phi_-^{(1)}= -{\mu\cosh{t\over\t_2}\sinh{\phi\over\t_2}\over 2\sqrt{2}\t_2^{3/2}(\cosh{\phi\over\t_2})^{5/2}}~.}
From \pfe, we see that $\Phi_-^{(1)}+\delta\Phi_-^{(1)}$ is indeed time-independent.

\listrefs
\end